\documentclass[]{interact}

\usepackage{epstopdf}
\usepackage[caption=false]{subfig}

\usepackage[numbers,sort&compress]{natbib}
\bibpunct[, ]{[}{]}{,}{n}{,}{,}
\renewcommand\bibfont{\fontsize{10}{12}\selectfont}

\theoremstyle{plain}

\theoremstyle{definition}

\theoremstyle{remark}

\usepackage{hyperref}
\usepackage{amsfonts}       
\usepackage{float}
\usepackage[utf8]{inputenc} 
\usepackage[T1]{fontenc}    
\usepackage{amsmath}
\usepackage{url}
\usepackage{xspace}
\usepackage{graphics}
\usepackage{multirow}
\usepackage{array}
\usepackage{breqn}
\usepackage{bbm}
\usepackage{enumitem}
\usepackage{wrapfig}
\usepackage{capt-of}
\usepackage{lineno}
\usepackage{mathtools, cuted}
\usepackage{mwe}
\newcommand{\xhdr}[1]{\vspace{0.1mm}\noindent{{\bf #1.}}}
\usepackage{xcolor,colortbl}
\usepackage[ruled,vlined]{algorithm2e}
\usepackage{textcomp}
\usepackage{booktabs}       
\usepackage{nicefrac}       
\usepackage{microtype}      

\usepackage{setspace}
\begin{document}

\title{Modeling the Heterogeneity in COVID-19's Reproductive Number and Its Impact on Predictive Scenarios}


\articletype{ORIGINAL RESEARCH ARTICLE}

\author{
\name{Claire Donnat\textsuperscript{a}\thanks{CONTACT Claire Donnat. Email: cdonnat@uchicago.edu}  and Susan Holmes\textsuperscript{b}}
\affil{ \textsuperscript{a}Department of Statistics, University of Chicago, 5747 S Ellis Ave, Chicago IL 60637, USA;  \textsuperscript{b}Department of Statistics, Stanford University,
   390 Jane Stanford Way, Stanford CA 94305, USA;}
}

\maketitle

\begin{abstract}

 The correct evaluation of the reproductive number $R$ for COVID-19 ---which characterizes the average number of secondary cases generated by each typical primary case--- is central in the quantification of the potential scope of the pandemic and the selection of an appropriate course of action. In most models, $R$ is modeled as a universal constant for the virus across outbreak clusters and individuals --- effectively averaging out the inherent variability  of the transmission process due to varying individual contact rates, population densities, demographics, or temporal factors amongst many. 
Yet, due to the exponential nature of epidemic growth, the error due to this simplification can be rapidly amplified and lead to inaccurate predictions and/or risk evaluation. From the statistical modeling perspective, the magnitude of this averaging's impact remains an open question: how can this intrinsic variability be percolated into epidemic models, and how can its impact on uncertainty quantification and predictive scenarios be better quantified? In this paper, we propose to study this question through a Bayesian perspective, creating a bridge between the agent-based and compartmental approaches commonly used in the literature. After deriving a Bayesian model that captures at scale the heterogeneity of a population and environmental conditions, we simulate the spread of the epidemic as well as the impact of different social distancing strategies, and highlight the strong impact of this added variability on the reported results. We base our discussion on both synthetic experiments --- thereby quantifying the reliability and the magnitude of the effects --- and real COVID-19 data. We emphasize that the contribution of this paper focuses on discussing the importance of the impact of $R$'s heterogeneity on uncertainty quantification from a statistical viewpoint, rather than developing new predictive models.
\end {abstract}


\section{Introduction}\label{sec:intro}

First detected in Wuhan (Hubei Province, China) in December 2019, the current COVID-19 pandemic has thrown the entire world in a state of turmoil, as governments closely monitor the spread of the virus and have taken unprecedented measures to contain and control outbreaks. 
In this context, the provision of accurate predictive scenarios is crucial for informing policy makers and deciding on the best course of action. Much attention has focused on the monitoring of one quantity: the pandemic's reproductive number $R$. This parameter is indeed key in  almost all contagion models, whether these scenarios are drawn using variants of the Susceptible-Exposed-Infected-Removed (SEIR) deterministic equations \cite{hethcote2000mathematics,kermack1927contribution,wu2020nowcasting,read2020novel} or of exponential growth models \cite{zhao2020preliminary}.
\\

\subsection{Reproductive Number(s)}
By definition, the reproductive number $R$ characterizes the expected number of secondary cases caused by an infectious patient.  Experts usually contrast the  {\it basic} reproductive number (typically denoted $R_0$) --- which assumes that the population is completely susceptible and is well adapted to the modeling of a completely novel virus --- with the context-dependent \textit{effective} $R_t$ --- which assumes a mixed population of susceptible and immune hosts at time $t$, and which varies with time and the implementation of various policies. Regardless of these assumptions, another way of understanding these reproductive numbers (denoted more generally as $R$ throughout this paper) is through their decomposition as the product of three terms \cite{daley2001epidemic}:
\begin{equation} \label{eq:r0}
    R = \tau \bar{c} D_I
\end{equation}
where $\tau$ is the transmissibility (i.e., probability of infection given a contact between a susceptible and an infected individual), $\bar{c}$ is the average number of contacts per day between susceptible and infected
individuals, and $D_I$ is the duration of infectiousness --- that is, the number of days during which an infected patient can be expected to contaminate others. $R$ thus serves as an epidemiological metric to describe the propensity of the epidemic to grow: the outbreak is expected to propagate if $R$ is greater than $1$, or to naturally subside if $R$ is strictly less than 1. As recently highlighted by Delameter et al. \cite{delamater2019complexity}, this coefficient inherently depends on some individual and local population characteristics, as well as seasonal and environmental variables. In particular, as shown in  Eq. \ref{eq:r0}, $R$ is intrinsically tied to temporal and spatially-varying factors, such as a population's age demographics and density, political or environmental variables, cultural or social dynamics --- all favoring or diminishing the rate of contacts $\bar{c}$ between individuals. This decomposition thus brings to light several sources of variability for $R$:
\begin{itemize}
    \item \textit{Temporal variability:} as time progresses and public policies (e.g, mask wearing, social distancing, etc.) change, we expect the contact rate, as well as the transmissibility to vary --- thereby  introducing a change in the expected number of secondary cases that $R$ models. 
    \item \textit{Subject variability:} communities can be well modeled by social networks, in which edges represent the contacts patterns. These models capture the important heterogeneity in the population's contact rates, typically correlated to one's age, profession, etc. 
\end{itemize}

These sources of variability pose a serious challenge to epidemiological models, particularly in light of the increasing  accounts of super-spreading phenomena in the literature \cite{cave2020covid,gomez2020mapping,hu2020identification,read2020choir, zhang2020evaluating}. Evidence seems to hint to the Pareto nature of the reproductive number, with 10\% of the individual cases potentially accounting for almost 80\% of the virus spread \cite{mackenzie}: the current pandemic appears to be driven by rare, yet important contagion events. Scientific evidence thus points towards the huge variability in the distribution of $R$, which should be considered as a random and potentially heavily skewed variable rather than a fixed number.

\subsection{Subject variability in epidemics models}

Current  epidemics models can generally be placed on either of the two ends of a spectrum, depending on their ability to account for subject and environmental heterogeneity. Indeed, while temporal variability is typically modeled using an effective time-varying $R_t$ instead of the constant $R_0$,  other sources of heterogeneity are seldom really accounted for by compartmental models, but can be included in agent-based models --- often at huge computational expenses. \\

\xhdr{Compartmental Models} Compartmental models for disease modeling are one of the most popular frameworks for predicting the evolution of an epidemic. These models build upon a division of the population into states or ``compartments'', the evolution of which being determined by a set of fixed, deterministic equations. In the case of COVID-19, the Susceptible-Exposed-Infected-Removed (SEIR) model seems to have been the most used among experts \cite{chatterjee2020healthcare,grant2020dynamics,he2020seir,pandey2020seir,wu2020nowcasting,zhao2020modeling}.  In this setting, each group (or community)  $k$ of size $N_k$ is split in one of four different compartments 
: people are either susceptible, exposed, infected  or removed (including recoveries and deaths). In this class of models, broadly speaking, the evolution of the populations in each compartment is modeled through versions of the following set of differential equations:

\begin{equation}
    \begin{split}
       \text{\textbf{S}usceptible:} \hspace{1cm}    \frac{d S_k(t)}{dt} &= -\frac{S_k(t)}{N_k} \frac{R^{(k)}_0}{D_I}I_k(t)  \\
           \text{\textbf{E}xposed:} \hspace{1cm}      \frac{d E_k(t)}{dt} &= \frac{S_k(t)}{N_k} \frac{R^{(k)}_0}{D_I}I_k(t)  -\frac{E_k(t)}{D_E} \\
         \text{\textbf{I}nfected:} \hspace{1cm}        \frac{d I_k(t)}{dt} &= \frac{E_k(t)}{D_E} -\frac{I_k(t)}{D_I}\\
                  \text{\textbf{R}ecovered:} \hspace{1cm}        \frac{d R_k(t)}{dt} &= \frac{I_k(t)}{D_I}\\
    \end{split}
\end{equation}

where:
\begin{itemize}
        \item $S_k(t), E_k(t), I_k(t)$, and $R_k(t)$ are the numbers of
susceptible, latent, infectious, and removed individuals
at time $t$ in group $k$;
\item $D_E$ and $D_I$ are the mean latent (assumed to be
the same as incubation) and infectious period (equal to
the serial interval  -- that is, the average time interval between the onset of symptoms in a primary and a secondary case \cite{rai2020estimates} ---  minus the mean latent period); 
\item $R^{(k)}_0$ is the basic reproductive number in population $k$.
\end{itemize}

In these models, the parameters $D_E$ and $D_I$ are typically fixed and taken from medical reports, while $R_0^{(k)}$is inferred and fitted on the available data.
While simple from both a theoretical and computational perspective, this class of models exhibits nonetheless several drawbacks. First of all, this deterministic set of equations does not provide any natural uncertainty quantification --- a crucial aspect of any model, especially given that all the parameters that are fed into these equations are (informed) guesses, that come with their own level of uncertainties. But the main drawback consists in these models' agnosticism to population heterogeneity. Indeed, while temporal transmissibility is captured by certain compartmental models which have replaced the $R_0$ with a context-dependent $R$ --- thereby allowing to capture the effect of public policies, seasonal effects, etc.---, the heterogeneity of the $R$ in the population is seldom considered. While some studies have introduced stochastic components in SEIR models (for instance in the study of Ebola \cite{lekone2006statistical}), it is however not standard to
 take into account any component of stochasticity in the reproductive number. As a way around this heterogeneity, some \cite{dolbeault2020heterogeneous,deforche2020age, lyra2020covid} have split compartments by stratifying with respect to age to account for varying contact rates across age groups. Yet, this stratification only takes into account one axis of variability (e.g., the age), and neglects other sources of variance, such as people's occupations, varying risk-exposure indices, etc. It is thus insufficient to reproduce the variability that is observed in real life, and thus potentially hinders the accuracy of any downstream analysis and predictive scenarios.   
 
 In fact, the  ``universal'' $R$ used in epidemiological models  to characterize the disease can be thought of as a general summary statistic, averaged over individuals and populations --- thus discarding any form of local variability. In the standard framework, the heterogeneity of $R$ is partially accounted for by fitting the model to local data (typically at the country or county level), but thus effectively (a) discarding information that could be shared across groups by fitting each local model independently and (b) neglecting the inherent variability in this coefficient at the individual level. The latter can in particular become problematic, due to the exponential nature of the epidemics growth --- a phenomenon that we propose to investigate here. In other words, this class of model typically works with a granularity at the group level (age group, etc.). Thus, while this framework can capture global tendencies across individuals (social mores, policy measures, etc), it is difficultly amenable to incorporating sufficient heterogeneity (due to subject characteristics, environmental variables, etc.) in the reproductive number.\\
 
\begin{figure}
    \centering
    \includegraphics[width=10cm]{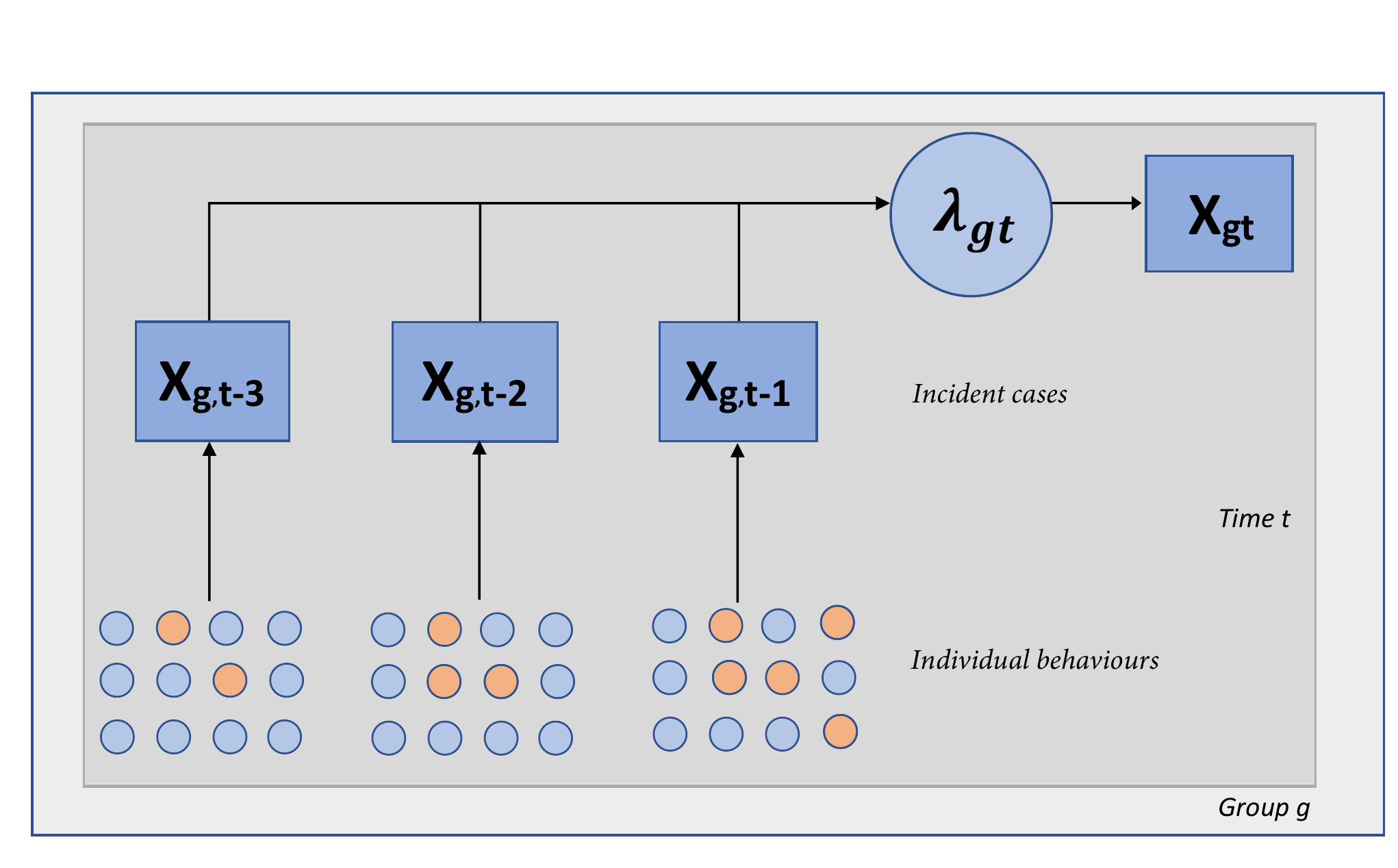}
    \caption{Plate model for simulating the transmission dynamics in agent-based models. At every time step, individual behaviours are simulated to infer the number of contacts and new infections.}
    \label{fig:plate_agent}
\end{figure}

 \noindent \xhdr{Agent-based Models}  On the other extreme side of the spectrum, agent-based models allow maximal flexibility by modeling the behavior of each single agent \cite{akbarpour2020socioeconomic,chang2020modelling,kai2020universal,rockett2020revealing,silva2020covid}. In \cite{akbarpour2020socioeconomic} and \cite{chang2020modelling} for instance, the authors are able to leverage mobility data (typically acquired using cellphone data) in order to simulate interactions and transmission dynamics, and consequently, to analyze the impact of different social distancing policies. Compared to compartmental models, the granularity of this class of models is at the individual level: each agent's trajectory is computed in order to infer the disease's transmission dynamics (see Fig. \ref{fig:plate_agent}). While able to capture a wide variability of individual behaviors, the success of such models is contingent on (a) huge computational resources to run the simulations and (b) large, quality datasets on mobility and community dynamics on which to base behavioral estimates. Moreover, these models make it more difficult to account for the effect of exterior variables, such as weather conditions, impacting  all agents. Thus, while this class of models allows for maximal subject heterogeneity, it does not lend itself to the modeling of global tendencies across agents.  \\
 
 Given the limitations and cost of estimating behaviours at the agent-level, in this paper, we propose to bridge these two classes of models through a Bayesian perspective.  Our approach allows us to capture the effect of the variance in the reproductive number on predictive scenarios, whilst remaining computationally tractable. The main purpose of this paper is to understand the impact of this heterogeneity on potential downstream epidemic scenarios. As such, our focus will be on quantifying the impact of the heterogeneity on the distribution of projected case estimates and worst-case scenarios. In the first section, we begin our discussion with a set of synthetic experiments, which allow us to quantify the discrepancy between the standard ``averaged'' and our ``distributional'' reproductive numbers. In the second, we present a Bayesian model that we have found to be amenable to the modeling of COVID-19 trajectories observed in real life. We discuss the differences that it yields compared to more standard SEIR models in the third, and its potential implications for policy makers.\\

\noindent \xhdr{Additional note} We emphasize that the randomness in the $R$ that is being considered here is solely due to population heterogeneity and potential environmental variables,  but not to measurement error. In other words, we assume that the variability might not be due to testing capacities and under-reporting. Our Bayesian formalism could be extended to include this other level of randomness, but for the purpose of this particular discussion, we focus on subject heterogeneity, and leave the inclusion of measurement error to later improvements. We discuss this particular point in greater depth at the end of this paper.

\section{The impact of the added variability}\label{sec:sim}

Intuitively, the simplification of the distribution of $R$s in a population to a constant value can be justified by the assumption that the dynamics of the pandemic are similarly described by the trajectory estimated using the average $R$, or the average of the epidemic's trajectories with varying $R$. Yet, because the number of new cases each day depends exponentially on the history of the trajectory, this averaging approximation  might come at a huge accuracy cost in prediction models. In this section, we aim to provide a more quantitative description of the potential effects of this additional variability on the model. Let us start with two naive experiments.\\ 

\xhdr{First Experiment: Inherent effect of the randomness on the model} In the first experiment, we consider a simplified epidemic exponential growth model. Given the value for the reproductive number $R$, each new infectious case $i$ generates a Poisson($R_i$) number of new cases the following day. This simplified model amounts to considering an instantaneous incubation period and a duration of infection of only one day. We can model the variability in people's secondary infection rates by considering the secondary infections as Poisson-generated counts. At each time $t$, the number of new incident cases the next day is thus generated as:
$$X_{t+1}  = \text{Poisson}(\lambda_t), \quad  \mathbb{E}[\lambda_t]  =  {X_t} \mathbb{E}[R]  $$
where $\lambda_t$ is the total incidence rate on day $t$.
Let us first study the  heterogeneity in reproductive numbers by comparing the baseline model ($M_0$) in which $R=R_0$ is constant, to one where $R$ is a random variable ($M_a$) centered around the same value $R_0$:
$$M_0:\quad X_{t+1}  = \text{Poisson}(X_t R_0) \hspace{0.5cm} \text{vs} \hspace{0.5cm}  M_a:\quad X_{t+1}  = \text{Poisson}(\lambda_t ), \quad \mathbb{E}[\lambda_t] = X_tR_0  $$
where $\lambda_t$ is the total aggregated reproductive number for day $t$, and is a function of the number of cases $X_t$ at day $t$.
To understand the effect of the variance in $\lambda$, we consider three probability distributions for $\lambda_t$, yielding different coefficients of variation $CV$ (defined as the ratio of standard deviation over mean, and which we use here as a metric to characterize the dispersion of the distribution of $\lambda_t$):
\begin{itemize}
   \item \underline{\textbf{$M_0$: Constant $R$, $CV(\lambda_t) = 0 $}:} in this ``null'' model, we assume that $R$ is a fixed quantity across subjects. The aggregated incidence rate $\lambda_t = \sum_{i=1}^{X_t} R_{i}^{(t)}$ for the group of people infected at day $t$ is given by:  $R_t = X_t R_0$.
    \item \underline{\textbf{$M_1$: Variable $R$, $\lim_{X_t \to \infty} CV(\lambda_t) = 0$}}, with $\lambda_t \overset{\Delta}{=} \Gamma(\alpha R_0  X_t,\alpha)$.  This first alternative model is equivalent to considering that each infected case at day $t$ has a random emission rate sampled from a Gamma distribution $R_i \overset{\Delta}{=}  \Gamma( \alpha R_0, \alpha)$, so that: $\lambda_t = \sum_{i=1}^{X_t} R_i \overset{\Delta}{=}  \Gamma( X_t \alpha R_0, \alpha)$. In this scenario, on average, the aggregated incidence rate $\lambda_t$ for day $t$ is the same as in model $M_0$, but the variance of the model is given by $\text{Var}[\lambda_t] = \frac{X_t R_0}{\alpha}$. Thus, the variance of the model scales linearly with the number of cases, while its coefficient of variation is: $CV = \frac{1}{\sqrt{\alpha R_0 X_t}}$ and tends to $0$ as the number of cases increases.
    \item \underline{\textbf{$M_2$:  Variable $R$, Constant $CV(\lambda_t)$}}, with $\lambda_t \overset{\Delta}{=} \Gamma(\alpha R_0, \alpha/X_t)$. This is a multiplicative model that assumes that the generation rate $\lambda_t$ scales linearly with the number of cases $X_t$, such that: $\lambda_t \overset{\Delta}{=}  X_t \Gamma(\alpha R_0, \alpha ) \overset{\Delta}{=} \Gamma(\alpha  R_0, \alpha/X_t) $. In this scenario,  the model has variance $\text{Var}[\lambda_t] = \frac{X_t^2 R_0}{\alpha}$, and constant coefficient of variation  $CV = \frac{1}{\sqrt{\alpha R_0}}$.
\end{itemize}

 These configurations allow us to study models with different levels of dispersion around the expected mean $R_0 X_t$ --  which we use to quantify the amount of subject heterogeneity.
Using this very simple generative model and starting with 100 infections, we generate 5,000 different epidemic trajectories assuming fixed $R_0$ and variable $R$s.\\

\begin{figure}
    \centering
    \includegraphics[width=\textwidth]{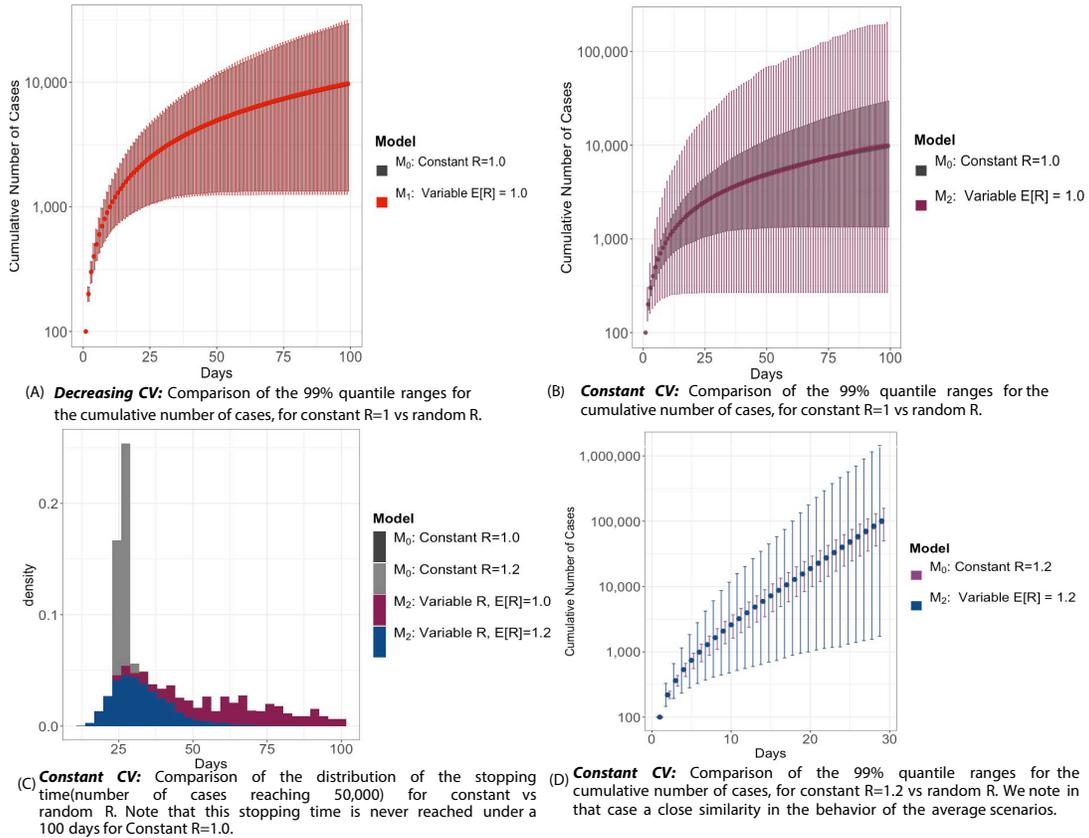}
    \caption{Simulations of the different epidemic trajectories, using various models for the daily new incident cases generation rate $\lambda_t$: a first benchmark model with constant, ``averaged'' $R_0$ (Model $M_0$, shown in grey shades on panels A, B and C and in purple on panel D), a model with variable $R$ and decreasing Coefficient of Variation (CV)  (Model $M_1$, with $CV \to 0$, shown in red on panel A), and a model with constant CV (Model $M_2$, shown in deep purple or blue on panels B , C and D). As a reminder, the coefficient of variation is defined as the ratio of the standard deviation over the mean, and can be used as a way of quantifying the dispersion of the model. We note in particular how seemingly alike the trajectories seem in average for $R_0=1$ in models $M_0$ and $M_2$ (panel B), but how substantially different their tail estimates are. }
    \label{fig:simulations}
\end{figure}

The results of these simulations are displayed in Fig. \ref{fig:simulations}. Based on those simulations, we make the following observations.
 
 \xhdr{(a) The inclusion of some amount of randomness alters the distribution of the trajectories, especially in the tails} Fig. \ref{fig:simulations}(A,B,D) show the $99$\% quantile ranges of the trajectories at each time step, for a mean value of $R_0 \in \{1,1.2\}$, and where the variable $\lambda_t$ is sampled from models $M_0$, $M_1$ and $M_2$. These figures highlight that  some of the most striking differences between models are located in the tails. In particular, Fig. \ref{fig:simulations}(B) presents a choice of parameters for model $M_2$ for which, while the mean number of cases appears to be similar, the tails (represented by the $99^{th}$ quantile) differ by orders of magnitude. This is an important observation: average predictions for the fixed and variable $R$ models can look seemingly the same, yet their spread and catastrophic scenarios are radically different. This divergence particularly striking between models $M_0$ and $M_2$ and increases for $R_0=1.2$, a scenario in which the pandemic is expected to rapidly increase.
 
 \xhdr{(b) The higher the volatility of $R$, the greater the divergence with models assuming constant $R$} The tail discrepancy between models $M_0$ and $M_1$  --- whose coefficient of variation  tends to 0 as $X_t$ increases --- is less substantial than with model $M_2$ (with constant positive CV, independently of $X_t$) : the $99^{th}$ quantile for model $M_1$ slowly diverges from $M_0$, but the difference after 100 time steps is only of the order of $2$ percent (compared to orders of magnitude for $M_2$). While intuitive, this is important to note: the adequacy and impact of the averaging of the $R$ made by SEIR models are contingent on the $R$'s inherent subject variability in the population. The more volatile the distribution of $R$, the greater the potential deviation of the epidemic trajectory from the projected SEIR one.
 
 \xhdr{(c) Worst-case scenarios are different -- not only in magnitude, but also in the events that they allow} Let us consider a specific event, which we choose here to be ``The cumulative number of cases reaches 50,000'', and let us display in Fig.~\ref{fig:simulations}C the distribution of its associated stopping time: the histograms describe the distribution of the value of this stopping time, given that it was reached in less than 100 days. For $M_0$ with $R_0=1$, this stopping time $\tau = \min\{t \in \mathbb{N}: 
\sum_{s=1}^t X_s  \geq 50,000 \}$ is never reached. It is nonetheless reached in 3\% of cases using a varying $R$ (with model $M_2$, in Fig \ref{fig:simulations}C), thus making it a non-zero probability event and enlarging the space of possible events. The variable-R model thus presents a wider scope of worst-case scenarios than the ones predicted using a constant, average $R_0$ --- a potentially crucial fact whilst having to decide on any type of policy. 
 \\
 
\xhdr{Second Experiment: effect of the randomness on the estimation procedure} We have shown that a constant $R_0$ might lead to an incorrect model of the distribution of probable epidemic trajectories -- we now also assess how the error induced by averaging is also reverberated in the estimation procedure.  In this second experiment, we simulate an exponential growth of the number of incident cases over the course of 20 days using model $M_2$ and a gamma-distributed $R$ with shape $1.2$ and rate 1. This mimics a scenario under which $R$ varies every day, thus accounting for some temporal effects (weekend vs week days), subject-effects across newly infected cases, etc. Let us now try to recover the reproductive number $R$ using the Exponential Growth model in the {\tt R}-package {\tt R0}. The average difference between the recovered and true mean $R_0$ over 1,000 simulations is 2.94 (with only 8.5
\% coverage by the recovered confidence intervals). 
This brings to light two new observations: (a) standard $R$ estimation procedures --- which assume a constant fixed $R$ --- seem to perform even less well with variable $R$, and (b) the usual confidence intervals are too narrow, and do not correctly account for the high uncertainty of the predicted $R$ value.  \\

 In light of these synthetic experiments, assuming the reproductive number $R$ to be constant comes at a huge cost in terms of accuracy of the reported predictive scenarios. In particular, the worst-case scenarios associated to these predictions could be either (i) too optimistic without appropriately characterizing their uncertainty, (ii) unable to account for the existence of ``super-spreaders'' in the general population, and (iii) fail to allow certain rare events leading to the formation of outbreaks --- thus potentially misleading policy makers and begging the question: for the analysis of real data, how much variability do we need to account for in the modeling of $R$? 


From a statistical viewpoint, accounting for the $R$'s variability is similar to using a ``random-effects'' model --- that is, endowing the reproductive number $R$ with a distribution and allowing it to vary across subjects. In this paper, we assume that the distribution of $R$ remains stationary over time. We fit the model over periods of time where policies are expected to be similar, and where we do not expect a dramatic change in the distribution of the reproductive number. Further extensions of this model would include adding other independent variables.
In particular, a more granular estimation of the dependency of $R$ on geographical, weekday, weather, and other sources of information could make day-to-day variations in the $R$ provide more realistic epidemiological predictions of the outbreak propagation speed, as well as the expected times before hospitals reach capacity --- both crucial quantities for informing policy makers as they arbitrate between different courses of action, especially as drastic public health measures typically come at significant social and economical costs. We do not focus on fitting any of these predictors here, nor we will fit a temporal trend line to our estimation of $R$. We justify these simplifying assumptions and discuss potential extensions of model in the conclusion and discussion section of this paper.
We emphasize that our goal is not to come up with a new model or definition for $R$, nor to pretend to a better predictive model than experts in epidemiology. Rather, our focus is simply to assess -- as statisticians -- the effect of this added variability in predictive scenarios, in order to better understand how this variability is propagated in downstream analyses.

One of the hypotheses that we would like to test is if the heterogeneity of the $R$ coefficient can severely impact predictive scenarios for the outbreaks: how certain are we of the predictions that we are making? In light of the observed heterogeneity of the $R$'s, how confident are we of the transferability of a given policy in one country to another? 
 Here, we deal with stochasticity and limited/missing data using a Bayesian perspective. We begin by describing the Bayesian hierarchical model that we use to estimate the varying reproductive number $R$. This approach provides a more natural framework for uncertainty quantification through the provision of credible intervals. We show the impact of this variability on the predictive scenarios and the effect of public policy measures (e.g. social distancing or alternating lockdown days) that can be drawn using these models. All of our experiments here are  deployed on the current COVID-19 pandemic. The code and data used for this analysis are openly available on the authors' Github\footnote{Code and data at: \url{https://github.com/donnate/heterogeneity_R0}}.

\section{Model and Theory}\label{sec:model}


In this section, we begin by designing and fitting a Bayesian model to real COVID-19 data.  This model can then be used to evaluate the effects of different policies on outcomes of interest (daily and cumulative number of cases).\\

\xhdr{Model} 
Our model is similar to the one previously used in the experimental section of this paper and based on the non-parametric model developed by Fraser \cite{fraser2007estimating} and later used for estimating the $R_0$  in Cori et al \cite{cori2013new}. This model is well-established and implemented in the {\tt R}-package {\tt EarlyR} \cite{earlyR}, and it  has been used
in recent studies \cite{zhao2020preliminary}
to infer COVID 19's $R_0$. Instead of explicitly modelling the exposed and infected periods separately, this model foregoes the modelling of latent cases and relies solely on inferring the number of new cases from previous observations using an ``infectivity profile'' \cite{cori2013new}. In this setting, each infected case is expected to contaminate on average of $R_0$ patients (by definition) --- but the distribution of this number of new infections over the infectious period is given by a probability distribution which only depends on the time $s$ elapsed since infection. One could indeed imagine a patient becoming increasingly contagious over the first few days of the infection as the viral load builds up, and decreasingly so after the peak of the illness. This infectious profile is typically modelled as the quantiles from a gamma distribution. Since this quantity is generally unknown and hard to estimate from available data, Cori et al \cite{cori2013new} propose the use of the parameters of the serial interval (for which we typically have much more substantial observational data and means of estimation) as a good proxy.  


We call $X_t$ the number of new infectious cases each day, and let $I_t$ be the number of total new infections caused by the $X_t$ subjects infected on day $t$ (in short, $I_t$ is the total number of secondary cases due to the $X_t$ new cases on day $t$). The incidence on day $t$ conditioned on the history of previously infected cases can be modelled by a Poisson distribution of the form: 
$$ \forall t\leq T, \quad  X_t  \sim \text{Poisson} ( \sum_{s=1}^{t-1} w_s I_{t-s})  \quad$$
with: \begin{equation}\label{eq:condition}
    \mathbb{E}[I_t] = X_{t-s} \mathbb{E}[R_t].
\end{equation} 

This condition captures the fact that  $I_t$ is the sum of all the new cases generated by each single patient.   Moreover, the Poisson distribution is well suited to the study large populations in which we expect the size of the epidemic to remain small, thus allowing us to neglect finite population effects as a first approximation.  In the previous equation, $w_s = \frac{K-s + 1}{\frac{K+1}{2}}$ is a vector such that $\sum_s w_s=1$. Note that, contrary to Cori et al \cite{cori2013new},we have not taken $w_s$ to be the estimated serial interval. As detailed in Appendix \ref{appendix:dets}, our choice of $w_s$ provides indeed a better fit to the data in this Bayesian pipeline.

The crux of the problem consists in specifying an adequate distribution for $I_t$, such that it verifies the condition $\mathbb{E}[I_t] = X_{t} \mathbb{E}[R_t]$. We take $I_t$ here to be gamma-distributed, with shape and rate parameters  $a$ and $b$. While different choices of parametrizations allow condition \ref{eq:condition} to be satisfied, we opt for one that allowing the best fit and  amount of stochasticity:
\begin{equation}
    \begin{split}
        \forall g, \quad \alpha_g & \sim \mathcal{N}(0,1)\\
        \beta_g & \sim \mathcal{N}(0,1)\\
        \forall t, g,\quad I_{t,g} & \sim \text{Gamma}( (1 + e^{\alpha_g}), (1 + e^{\beta_g})/{X_{t,g}})  \\
        \forall t, g, \quad X_{t,g} & \sim \text{Poisson} ( \sum_k w_k I_{t-k} )\\
    \end{split}
\end{equation}\label{eq:model}
where the subscript $g$ denotes the group and emphasises the fact that each region/group is fitted independently and where we have taken the shape $a$ to have the form $1 + e^{\alpha}$, and $b$ to be parametrised as $1 + e^{\beta}$.
The full model is summarised by the plate model provided in Figure \ref{fig:plate}. We also provide in the appendix a more detailed discussion on the choice of the model, the choice of the priors, as well as all technical details related to the fitting procedure, and focus in the main text on the analysis of the subsequent results.

\xhdr{Interpretation of the model} This model ($M_1$) is akin to a multiplicative model, rather than an additive one ($\tilde{M}_1$, here provided for comparison):
$$M_1: \lambda_t  \overset{\Delta}{=} X_t \Gamma(a,b) \overset{\Delta}{=} \Gamma(a,b/X_t)  \hspace{1cm}  \text{vs} \hspace{1cm}  \tilde{M}_1: \lambda_t \overset{\Delta}{=}  \sum_{i=1}^{X_t} \Gamma(a,b) \overset{\Delta}{=} \Gamma(a X_t,b)$$ with $a = e^{\alpha}+1$ and $b = e^{\beta}+1$. While intuitive, model $\tilde{M}_1$ has a small coefficient of variation $CV = \sqrt{\frac{b}{aX_t }}$ (as discussed in section \ref{sec:sim}), and empirically fails to capture the important amount of stochasticity observed in the real data. We refer the reader to the Appendix for a more detailed discussion on model fit.
 By comparison, the selected model $M_1$  assumes a multiplicative effect of $X_t$ over $R$, $I_t =  \bar{R}X_t$ and yields a constant coefficient of variation ($CV = a^{-1/2}$) and standard deviation scaling linearly with the number of cases.
        This type of model (as shown in the appendix) provides a better fit to the data, which could be due to the fact that it allows greater dispersion and to also allow to account for under-reporting and/or asymptomatic cases contributing to the new incident cases through this multiplication. 

\begin{figure}
    \centering
    \includegraphics[width=10cm]{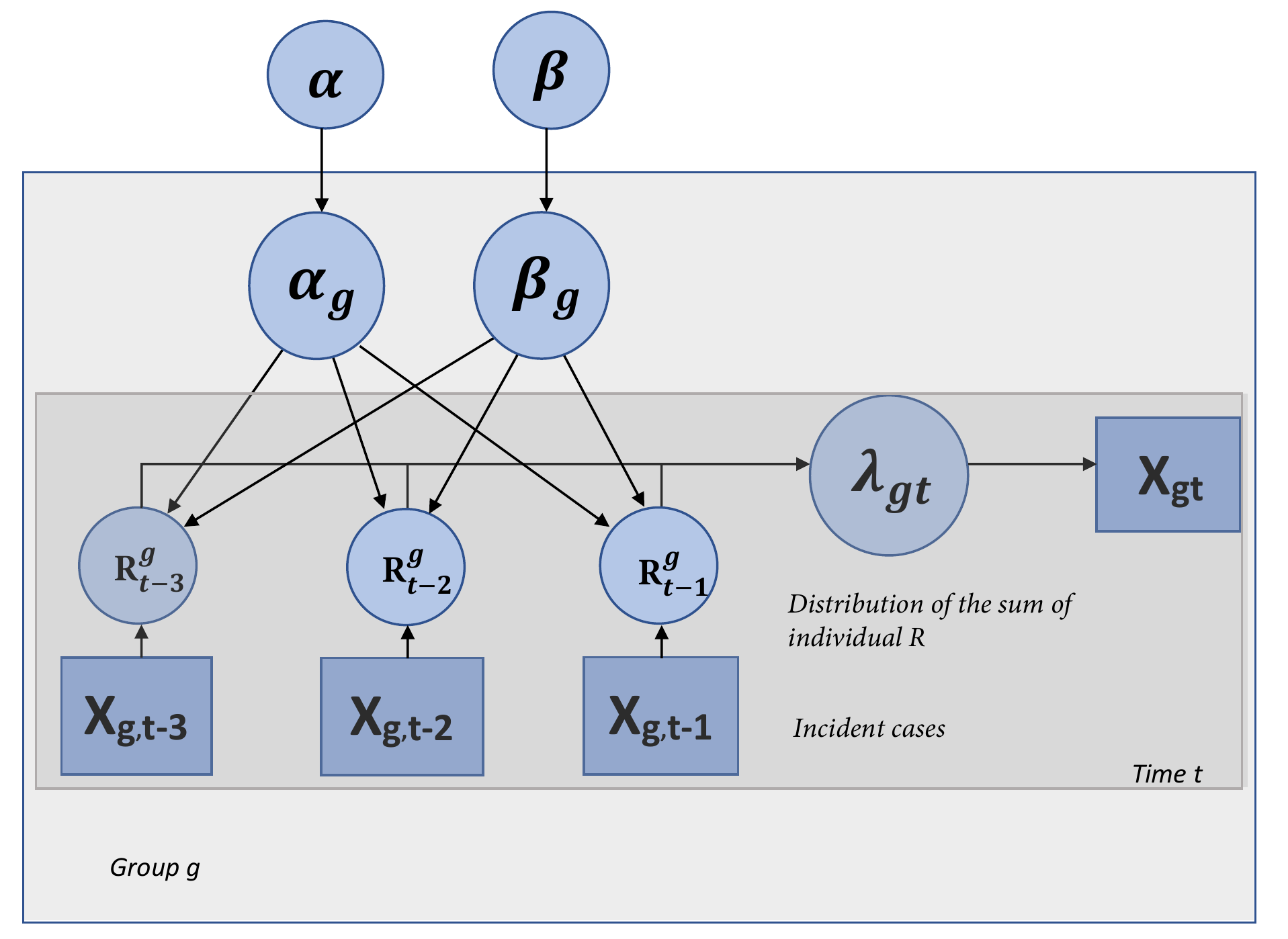}
    \caption{Plate model for the Bayesian Model described in Eq.\ref{eq:model}}
    \label{fig:plate}
\end{figure}

\xhdr{Fitting real data: Results} We fit the previous model on 14 different areas across the world, chosen arbitrarily by the authors but for which we expect diversity in policies, environmental variables and social mores. To represent Europe,  we selected France, Germany, Italy, Spain, Sweden, Estonia and the United Kingdom. The United States of America were arbitrarily chosen to be represented through three American states (California, Florida, and Texas). We also selected two South American countries (Mexico and Colombia), as well as South Korea and Russia. We use publicly available datasets on the number of incident cases that we preprocess and smooth using 7-day rolling averages (see Appendix \ref{appendix:dets} for further details). For each of these fourteen regions, we split the data and fit the model independently on seven time ranges consisting of the 30 consecutive days from March 15th, to October 14th 2020. This splitting allows us to train the model independently on shorter periods --- thus allowing us to control for policy changes or weather shifts. The fitting of this Bayesian model was performed using Hamiltonian Monte Carlo\cite{betancourt2017conceptual,betancourt2015hamiltonian,hoffman2014no} with the {\tt Rstan} R-library\cite{carpenter2017stan}. We refer the reader to the appendix for further details on convergence diagnostics.

\begin{figure}
    \centering
    \includegraphics[width=\textwidth]{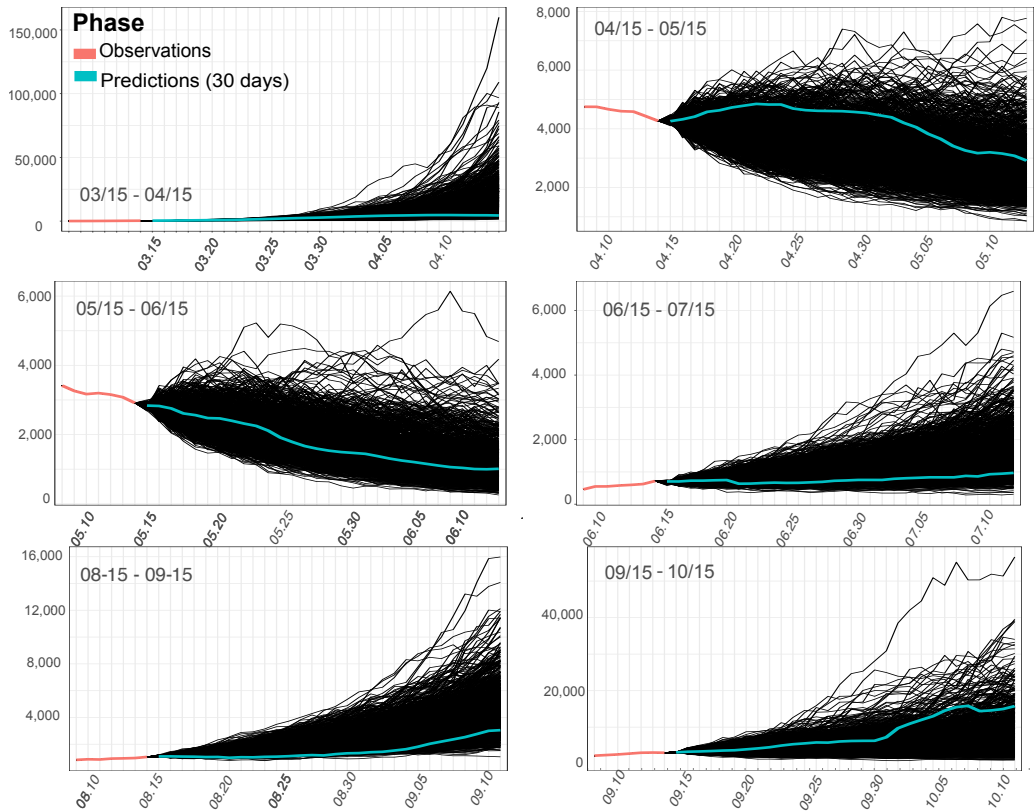}
    \caption{Comparison of estimated trajectories for the United Kingdom vs the actual ones. Here, the observation phase is in red, and the goal is to predict likely trajectories for the next 30 days, the observed one being indicated in teal).}
    \label{fig:trajectoriesUK}
\end{figure}

\begin{figure}
    \centering
    \includegraphics[width=\textwidth]{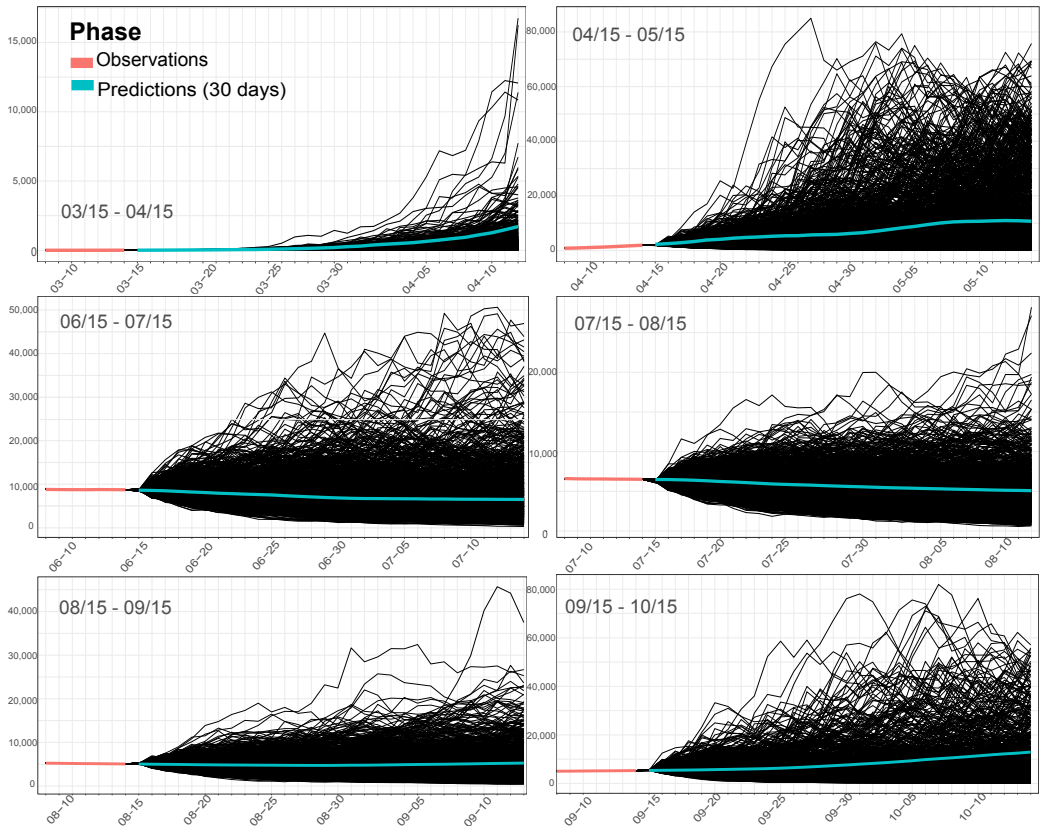}
    \caption{Comparison of estimated trajectories for Russia vs the actual ones. The observation phase is in red, and the goal is to predict likely trajectories for the next 30 days (the observed one being indicated in teal).}
    \label{fig:trajectoriesRussia}
\end{figure}

\xhdr{Assessing Goodness of Fit} Figures \ref{fig:trajectoriesUK} and \ref{fig:trajectoriesRussia} show  projected sample trajectories across time for two countries (United Kingdom and Russia) by way of illustration. The rest of the trajectories are provided as supplementary material on Github, along with the code\footnote{\url{https://github.com/donnate/heterogeneity_R0}} used to fit and generate these trajectories. 

\begin{figure}
    \centering
    \includegraphics[width=\textwidth]{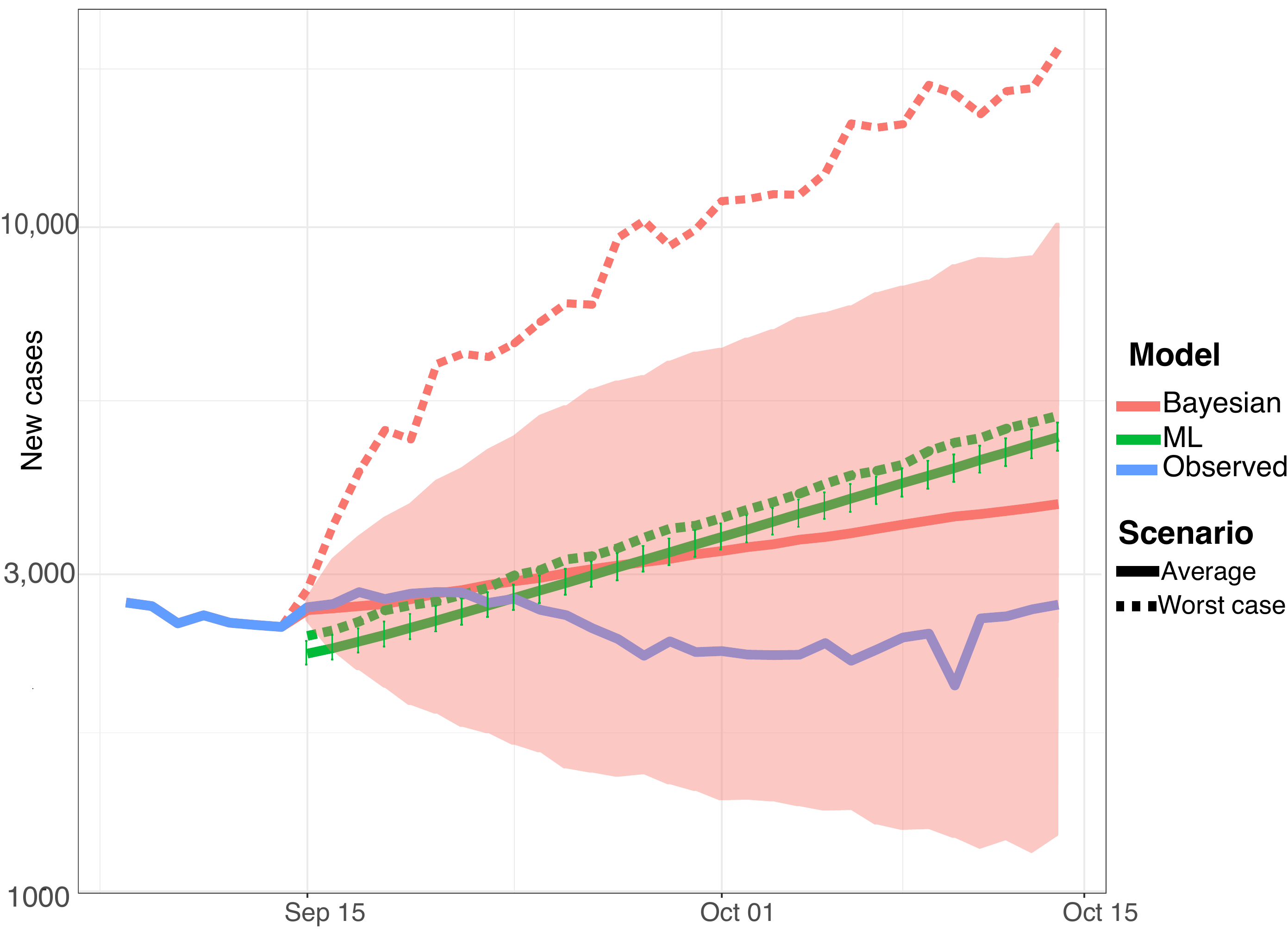}
    \caption{Comparison of the different predictive models for Florida. The light red color shows the $95^{th}$ quantile envelope associated to the Bayesian trajectories, while the green one is the envelope associated with the predictions using the reference {\tt R}-packages {\tt R0} and  {\tt projections}. The dashed line shows the worst case scenario associated with each of these methods, while the blue represents the actual observations.}
    \label{fig:comp_florida_summary}
\end{figure}

To create these figures, we generated Monte Carlo sample curves according to the model in Eq. \ref{eq:model} using seven days of observations (shown in red on the figure) to predict the next 30 (shown in teal), and the fitted parameters. At each sampling step, we reject new incidence numbers bigger than 1\% of the population. Our goal here is to assess the realism and goodness of fit of our curves compared to the actual observed curve. We note that the actual observed trajectories fit well within the bounds of what is predicted by our Monte Carlo samples --- thus indicating an agreement between our fitted model and the actual data and a good coverage of the trajectories.   To quantify model agreement further and benchmark it against existing approaches, we also compare our predictions and credible intervals with the confidence intervals predicted from the \texttt{R}-packages \texttt{R0}\cite{r0} and \texttt{earlyR}\cite{earlyR} packages, using their corresponding Maximum-Likelihood estimators on the same training data. The results are displayed in Table \ref{tab:EarlyR} for the period from mid-March to mid-April, and we show a comparison of the scenarios predicted by both models in Fig. \ref{fig:comp_florida_summary}. 

Overall, we note the consistency of the mean our estimates with the ones provided by the other $R$-packages. This is reassuring, since all models are based on the same type of model formalism. The main  difference lies in the width of the confidence intervals that the Bayesian method provides: as underlined by the second column in Table~\ref{tab:EarlyR}, the Bayesian model provides both larger credible intervals, from tens to hundreds-of-times wider that the confidence intervals provided for {\tt R0} or {\tt earlyR}. We also highlight that the confidence intervals provided by the Maximum Likelihood SEIR estimator implemented in {\tt R0} are in fact confidence intervals for a mean, constant $R$, whereas ours are for a distribution of $R$s. The confidence intervals for {\tt EarlyR} are closer, yet still substantially shorter than the credible intervals provided by the Bayesian method. As a consequence, our model yields better coverage of the observed trajectories (prediction 30 days ahead --- see an example for Florida in Fig. \ref{fig:comp_florida_summary}). In fact,  in 72.5\% of cases (across all time periods and all countries), the observed trajectory (predicted 30 days ahead) is fully contained by the envelope of sampled trajectories, compared to none for the {\tt EarlyR} predictions. We conclude that our model yields wider --- but more realistic --- estimates of the variability of the reproductive number and consequently, a more accurate distribution of plausible epidemic trajectories.  

\begin{table}[H]
\resizebox{\columnwidth}{!}{%
    \centering
    \begin{tabular}{|c|c|c|c|c|c|} \hline 
         Country & Bayesian Mean Estimate & {\tt $R_0$} ML- estimate & Ratio of Conf. Int. Width & {\tt earlyR} ML- estimate & Ratio of Conf. Int. Width\\
         &&&$ \frac{|CI|^{\text{Bayesian}}}{|CI|^{R_0}}$& &$ \frac{|CI|^{\text{Bayesian}}}{|CI|^{\text{earlyR}}}$     \\ \hline \hline
        Estonia & $1.31 \pm 1.29$ &$1.32 \pm 0.048$ & 26.9& 4.5 ${\pm 1.2 }$  & 1.1\\ \hline
        France & $1.37 \pm 1.37$ & $1.38 \pm 0.01$& 137&  1.8 ${\pm 0.08 }$  & 18 \\\hline
        Germany & $1.91 \pm 2.10$ & $1.51 \pm 0.005$ & 420& 2.1 ${\pm 0.07 }$  & 26  \\\hline
        Italy & $1.51 \pm 1.30$& $1.23 \pm 0.004$ &325 & NA &  NA \\\hline
        Russia &   $1.63 \pm 0.84$ & $2.16 \pm 0.0 42$ & 20& 2.2 ${\pm 0.82 }$  & 1.0\\\hline
        Spain &  $2.03 \pm 2.23$  &$0.99 \pm 0.01$&223& $2.5 {\pm 0.09}$  & 25  \\\hline
        Sweden &  $1.18 \pm 0.73$  &$1.44 \pm 0.02$ & 36.5& $1.7 {\pm  0.15} $ & 4.9 \\\hline
        UK &  $1.55 \pm 1.06$ & $1.67 \pm 0.01$ &58.9& $2.1 {\pm  0.10}$  & 10 \\\hline
        South Korea & $0.58 \pm 0.82$ & $0.96 \pm 0.018$ & 45.6 &NA  & NA \\\hline
        California & $1.49 \pm 1.20$ & $1.84 \pm 0.019$ & 63.2&$1.9 {\pm 0.28}$  & 4.3 \\\hline
        Texas &  $1.58 \pm 1.33$ & $1.98 \pm 0.03$ & 44.3& $2.1 {\pm 0.79 } $  &1.7\\\hline
    \end{tabular}
    }
    \caption{Comparison of the results across  a subset of regions for the first phase of the epidemic. The first column shows the mean (and 95\% confidence interval) associated to our estimated Bayesian $R$, while the second shows the estimate of the reproductive number obtained by Maximum-Likelihood estimation with the reference {\tt R-packages $R_0$}\cite{r0} and {\tt earlyR}\cite{earlyR}. The third column  provides an estimate of the ratio between the confidence bands associated to both methods. The NAs in the table correspond to entries where the {\tt EarlyR} algorithm has failed to produce any result. }
    \label{tab:EarlyR}
\end{table}

\begin{figure}
    \centering
    \includegraphics[width=\textwidth]{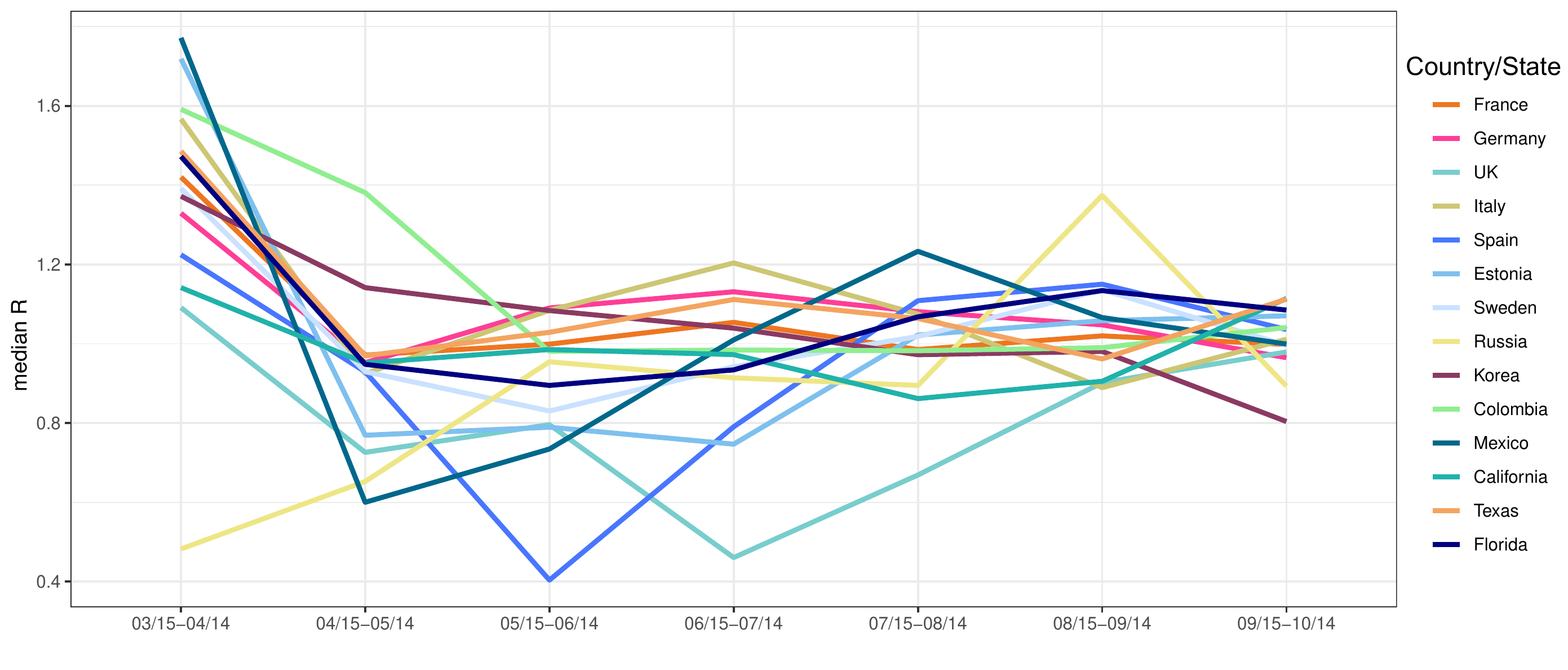}
    \caption{Trajectories of the median R (fitted on periods of 30 days, labeled on the x-axis) across countries/ states, as a function of time.}
    \label{fig:rhist}
\end{figure}

\begin{figure}
    \centering
    \includegraphics[width=\textwidth]{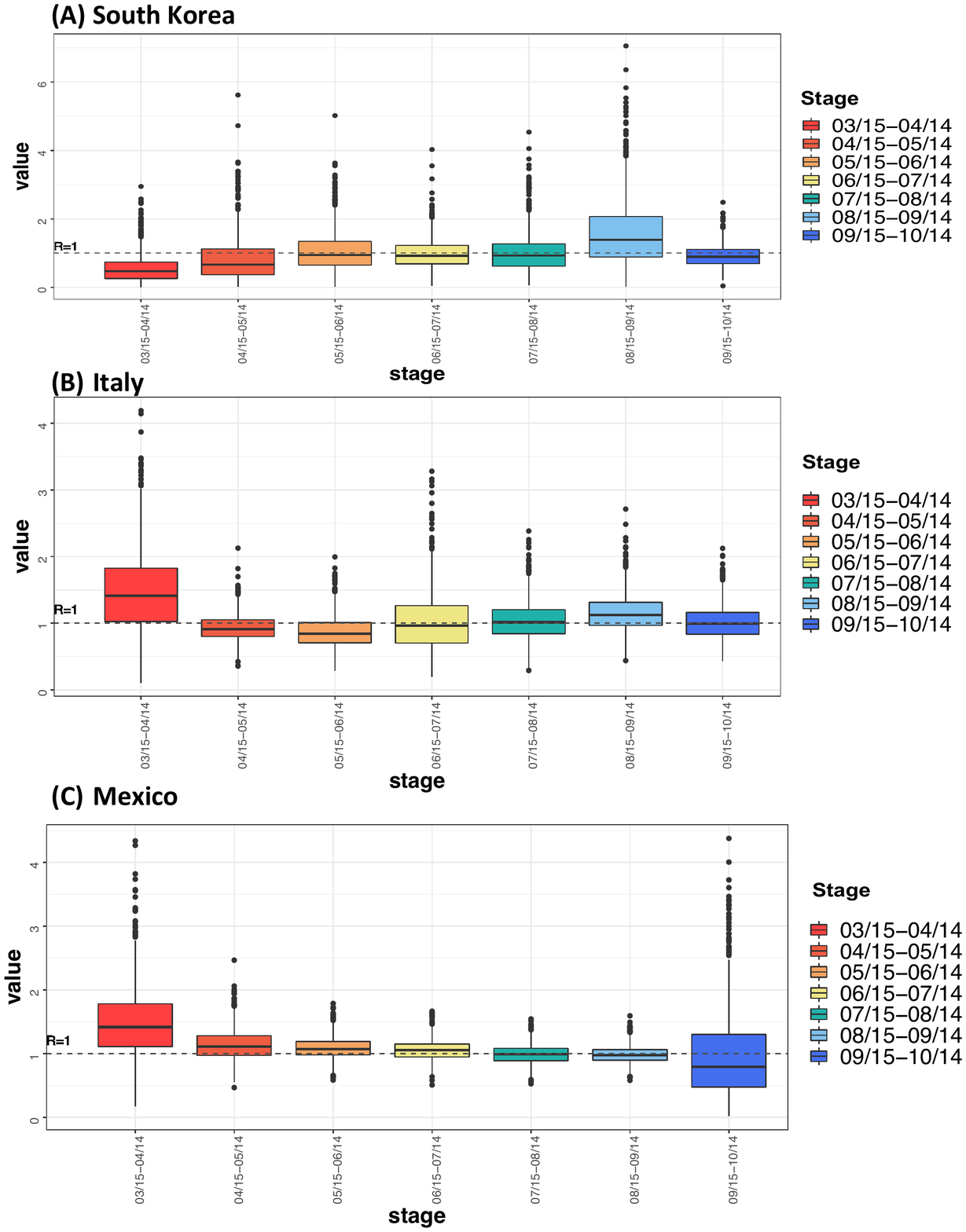}
    \caption{Boxplots showing the distribution of $R$ across three different countries and across phases. We note in particular the variability of the spread of the distribution across time.}
    \label{fig:rboxplot}
\end{figure}

\xhdr{Analysis and Interpretation of the results} Having established our model's goodness of fit, we turn to the interpretation of the fitted parameters. Figure \ref{fig:rhist} shows the distribution of the median of the fitted $R$ across countries and phases. We note, consistently to what we had been expecting, that the median $R$ dropped considerably after the first month (probably due to the strict lockdowns that were put in place roughly around mid-March in most of these regions), and has been plateauing since then around the threshold value of $1$ in almost all countries. This observation is consistent with the lingering of the epidemic that we are currently observing, and the relative stability of the pandemic in these different regions during the months of June until September. To enrich the discussion, instead of focusing solely on the median $R$, we provide a few examples of the boxplots of the distributions as a function of time in Fig. \ref{fig:rboxplot}. The last period of 30 days (mid-September to mid-October exhibits mostly plateauing behaviours and occasional rises in the 95\% quantile. As such, this particular time frame --- which could have acted as a precursor of the ``second wave'' of the pandemic, observed from November in many regions) --- does not highlight any significant uptake of activity of the pandemic in most countries. \\

We emphasise that the Bayesian model presented here provides a more complete portrait of the situations across countries than most SEIR-based models. Indeed, by considering a distribution, instead of the mean as a summary statistics, we are able to view interesting behaviours and spot nuances in countries' responses and handling of the pandemic. In particular, as shown in Table \ref{tab:table_quant}, all countries (but Korea, for which the epidemic has well under control mid-March) have managed to shrink the value of their average $R$ by an average of 0.55 points, yielding an average reduction of the $R$ to 65\% of its original value. But perhaps more striking is the average 1.1 reduction for these countries of the 95th quantile of the $R$ (66\% of its original value) --- highlighting the considerable reduction in the tail of the distribution. Not only do these numbers allow us to better quantify the effects of the different measures on mitigating the spread of the virus, it also allows to make more refined nuances between scenarios. For instance, for some countries (e.g Spain or Germany), the average $R$  has increased or remained comparable from  the summer period to the fall period, but we also note a sizeable reduction in the value of the 95th quantile: this seems to indicate a potential reduction of ``superspreader events''. This is potentially an encouraging sign for the handling of the pandemic, and thus, yields a more balanced view of the situation of the pandemic in these countries instead of the bleak picture provided by the sole examination of the behaviour of the mean. On the other hand, Texas for instance, has registered a 15\% increase in the value of its average $R$ between summer and fall, but a 200\% increase in its 95th quantile --- thus raising a warning signal for the potential resurgence of the pandemic in this particular region.
\begin{table}[]
    \centering
    \begin{tabular}{|p{3cm}||c|c|c|}
    \hline
     {\bf Mean (95th q.)} & {\bf 03-15-04/15} & {\bf 06/15-07/14 }& {\bf 09/15-10/14}  \\\hline
      California  & 1.5 ( 2.7)  &  {\color{green}1.1 (1.6) } & 1.0 ({\color{green}1.25}) \\\hline
 Colombia  &  1.4 (2.5) &   {\color{green}1.1 (1.5)} &  { 0.97} (1.3)  \\\hline
 Estonia  & 1.3 (3.1) &   {\color{green} 0.6 (1.5)} &  {\color{red} 1.0}({\color{red} 2.0})  \\\hline
 Florida  & 1.7 (3.2) &   {\color{green} 1.2 (1.7) }&  1.1 (1.6)  \\\hline
France  & 1.4 (2.7) &  {\color{green} 1.0} (2.6) &   1.1 ( {\color{green} 1.9})  \\\hline
Germany  & 1.9 (3.9) &   {\color{green} 0.8 (1.7)} &   {\color{red}1.1} ( {\color{green} 1.49}) \\\hline
Italy  & 1.5 (2.7)  &  {\color{green} 1.0 (1.9) }&  1.0 ({\color{green}1.5}) \\\hline
Korea  & 0.6 (1.4) &  {\color{red} 1.0 (1.8)} &   0.9 ({\color{green}1.6}) \\\hline
Mexico  & 1.4 (2.5) &  {\color{green} 1.0 (1.3)} &   0.9 ({\color{red}2.1})\\\hline
Russia  & 1.6 (2.4) & {\color{green} 0.9(1.2)}&   1.0 (1.3)  \\\hline
Spain  & 2.0 (4.3) &   {\color{green} 1.1 (2.1)} &   1.1 ({\color{green}1.7}) \\\hline
Sweden  & 1.2 (1.8) &  {\color{green} 1.0 } (1.7) &   1.1 (1.7) \\\hline
Texas  & 1.6 (2.9) &  {\color{green} 1.1 (1.5)} &   {\color{red} 1.3} ( {\color{red} 3.0}) \\\hline
UK  & 1.6 ( 2.6) &  {\color{green} 1.0 (1.8)} &  1.1 (1.7) \\\hline
    \end{tabular}
    \caption{Quantification of the evolution of the mean and $95^{th}$-quantile of the estimated distribution for $R$ across three different periods. Colors in the last two columns denote an increase (red) or decrease (green) of more than 0.2 points from the previous column. }
    \label{tab:table_quant}
\end{table}

\section{Evaluating the impact of adding heterogeneity in predictive scenarios.}\label{sec:predictions}

The second stage of our analysis consists in using our fitted model for the heterogeneous $R$ to predict the impact of different strategies on the outcome of the epidemic. 
Indeed, policy makers are currently faced with the difficult task of implementing efficient policies to limit the spread of the virus, while arbitrating between societal and economical costs. An inspection of the decomposition of the reproductive number provided in Eq. \ref{eq:r0} exhibits why a policy geared towards a lowering of the daily contact rate $\bar{c}$ should efficiently limit the spread of the virus. The goal of this section is thus to quantify the effect of governmental measures on slowing and mitigating the spread of the virus. Again, we emphasise that our study does not aspire to provide state-of-the-art prediction models, but rather to evaluate the effect of the additional variability on recommended measures. 

We model two types of interventions:
\begin{itemize}
    \item {\bf ones that act on the tails of the distribution }-- that is, interventions geared to mitigating the possibility of superspreader events (concerts, transportation, etc.). Such interventions are thus not targeted uniformly across the population, but rather at the highest quantiles of the distribution of $R$. Concretely, based on Eq. \ref{eq:r0}  and assuming that the variability in the transmissibility $\tau$ is less than the one associate to the contact rate $c$, these measures can be understood as capping the maximal contact rate. From the simulation perspective, we model these intervention by capping the values of the $R$ at different quantiles (the more severe the intervention, the lower the capping quantile).
    \item {\bf ones that are distributed uniformly on the population. } These  approaches are geared towards a reduction of the $R$ (or equivalently, the contact rate $c$) by measures applicable to the entire population, by shrinking each individual contact rate (e.g, mask wearing, generalized stay-at-home orders). From the simulation perspective, we model this by a reduction of the mean of the distribution or equivalently  (i.e, we multiply the shape parameter of the gamma distribution by a number less than one. The lower the multiplicative factor, the more severe the intervention).
\end{itemize}

In this section, we refer to the stringency (value of the quantile or of the multiplicative factor) as the ``level'' of the intervention. Note that actual real life interventions are often a combination of both effects, but we find convenient in this study of the impact of the heterogeneity onto the distribution to decouple the both.
All the results can be found in the supplementary materials on Github, and we provide in the main text illustration of these simulated scenarios for the case of France in the first month (03/15-04/14). 

\begin{figure}
        \centering
    \includegraphics[width=\textwidth]{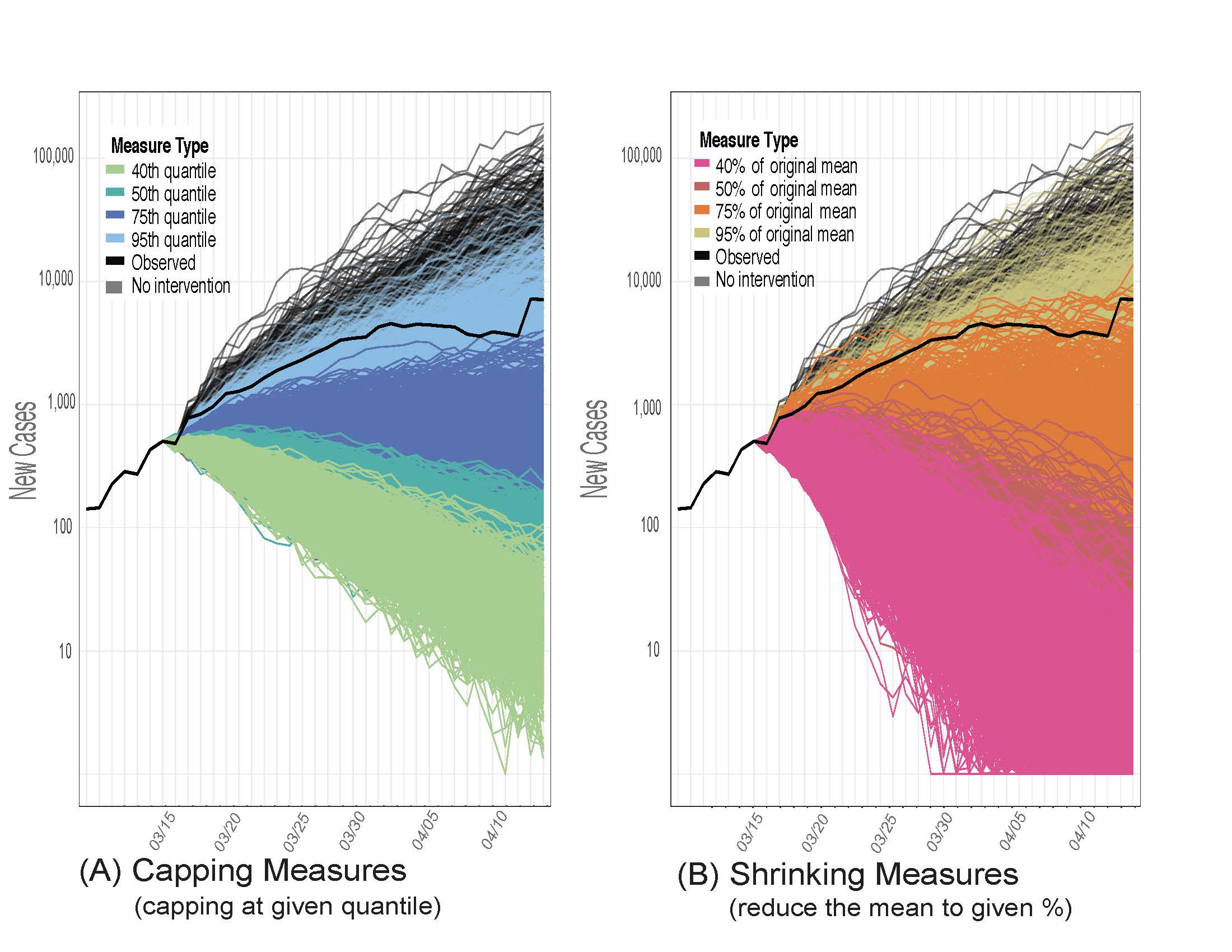}
    \caption{Comparison of the effect of the different types of interventions in France: ones acting on the tails of the distributions (panel A) and others distributed uniformly on the distribution (panel (B).}
    \label{fig:France1}
\end{figure}

Figure \ref{fig:France1}A shows the distribution of projected trajectories for measures aiming to cap the distribution of the $R$ at several pre-specified thresholds, whereas Fig. \ref{fig:France1}B shows the distribution of these trajectories for measures reducing the value of the $R$ as a distribution (mean shrinkage). In other words, we aim to compare here measures that are targeted at minimising the highest contact rates on the left (restrictions on the maximal gathering sizes, targeted lockdowns), to measures geared to minimising the entire distribution of contact rates (lockdowns, social distancing).  We note that both types of measures are efficient in reducing the spread of the epidemic (the more stringent the intervention, the faster and better the effect). In particular, a capping of the $R$ to its 50th quantile (or to 50\% of its original value) or less is efficient in making the epidemic recede. Measures shrinking the distribution as a whole also appear to have a faster effect: the epidemic receded within a month of an intervention reducing $R$ to 40\% of its original value (magenta curves on Fig. \ref{fig:France1}B), whereas it appears to take between one to two months   for an intervention capping $R$ to the 40th quantile of its original distribution (light green curves on Fig. \ref{fig:France1}A).

\begin{figure}
    \centering
     \includegraphics[width=\textwidth]{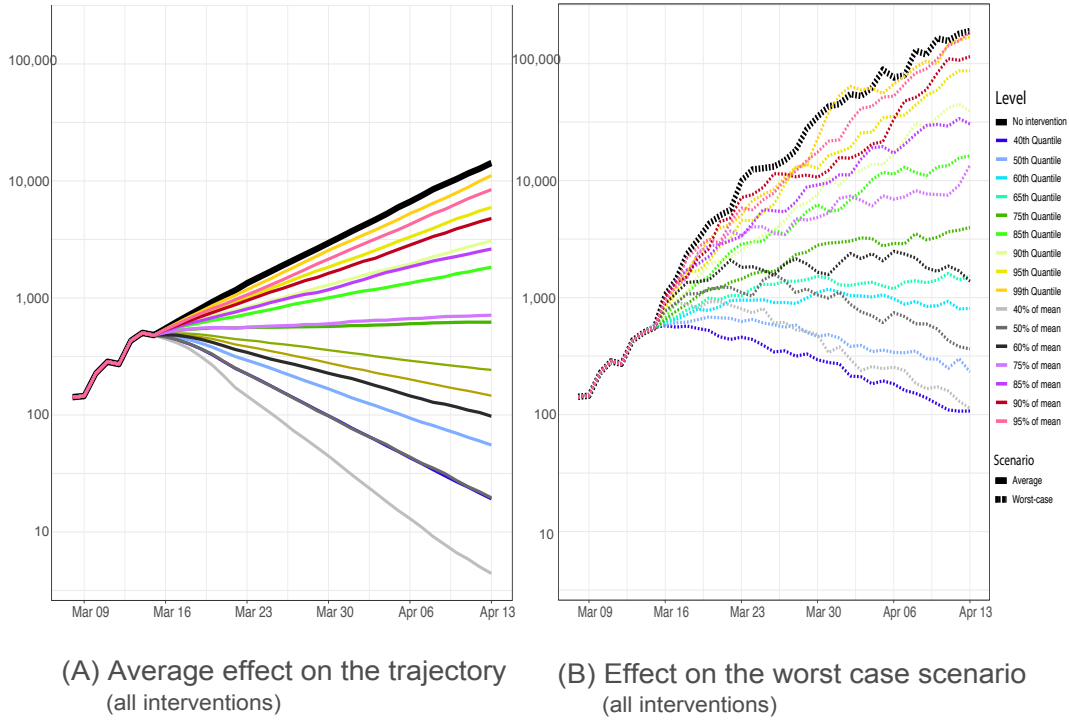}
    \caption{Comparisons of the different interventions in France on the distribution of the trajectories, represented by its average on panel A, and its maximum (worst case) on panel B. Interestingly, the average scenarios in panel A indicate that the two types of interventions yield comparable results (controlling for the level of stringency of the measure), whilst tail-focused measures --- in cooler colours--- yield better worst-case (95th quantiles) scenarios  than mean-orientated ones, here displayed with warmer colours (panel B).}
    \label{fig:France3}
\end{figure}

\begin{figure}
        \centering
    \includegraphics[width=\textwidth]{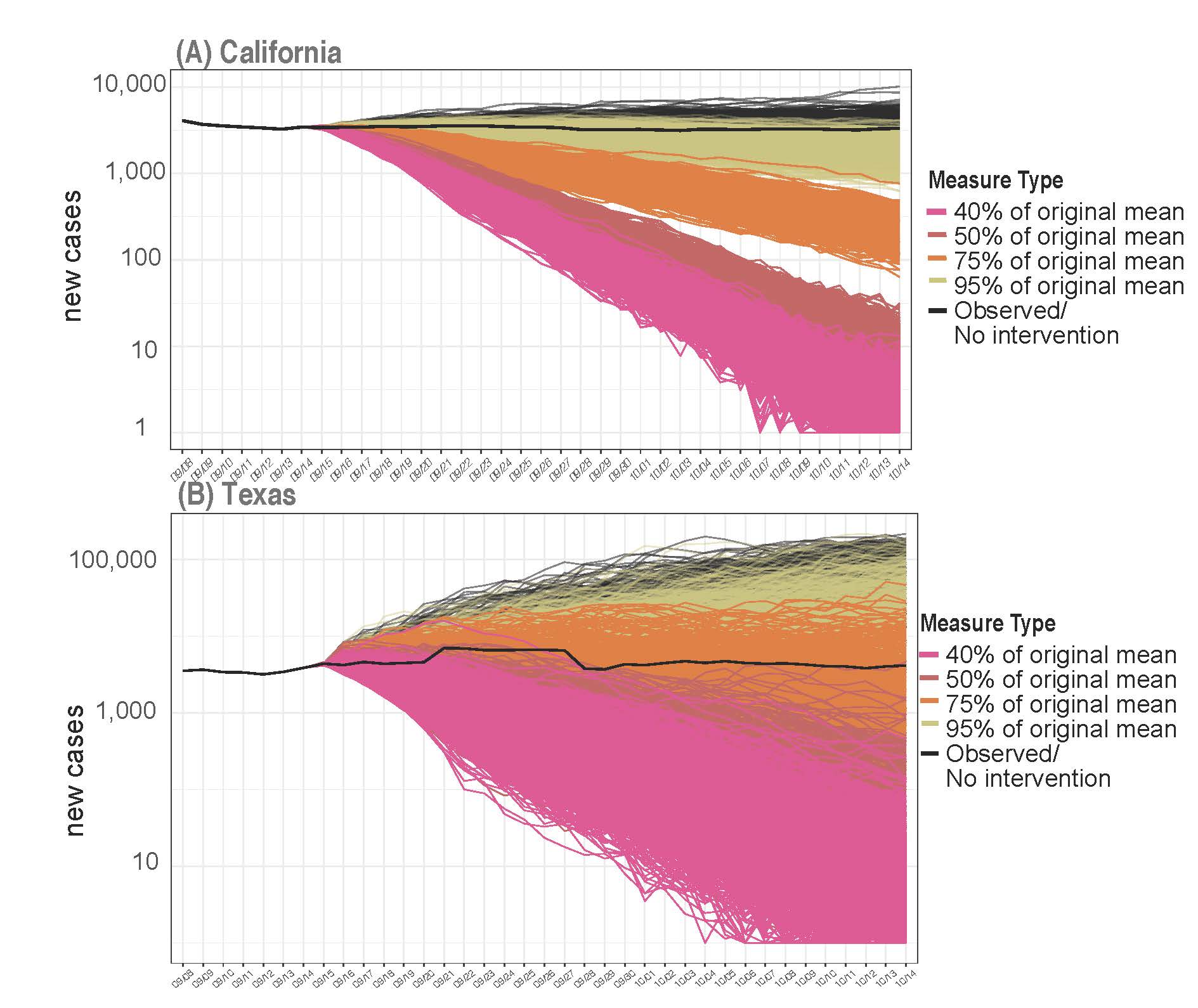}
    \caption{Comparison of the effect of the different types of interventions: California vs Texas, from 09/15 to 10/14. By Fig.\ref{fig:rhist}, both have similar median $R$ and average $R$, but the spread of  Texas is larger (see Table \ref{tab:table_quant}). This highlights the importance of the distribution of the reproductive number $R$ as a whole to determine the efficiency of any given measure.  }
    \label{fig:final}
\end{figure}

However, given the costs and difficult logistics of stringent measures acting on the whole distribution, it is important to quantify their optimality and efficiency with respect to ``tail interventions'', whose costs and burden on the population can potentially be lighter.  Fig.\ref{fig:France3} shows the average and 95th quantile (which we take as a measure of the ``worst-case scenarios) for the trajectories in France, while Table \ref{tab:final} quantifies the efficiency of the different measures across countries in the first month of the epidemic. Interestingly, the efficiency of the tail-oriented strategies are comparable to the ones reducing the entire distribution: the strategy consisting in capping $R$ to its 90th quantile already achieves a 78.5\% reduction in France --- more efficient than a strategy involving reducing the $R$ to 90\% of its original value. As shown in Table \ref{tab:final}, this effect of course widely varies across countries, some measures being more efficient in some countries than others. In Estonia for instance, a capping of the $R$ distribution at the 95th quantiles induces a reduction of 62.4\% of the epidemic, but reducing the mean to 95\% of its original value only induces a decrease of 34.4\%.  In Russia, the efficiency of these two measures is in fact reversed (34.8\% vs 47\% reduction). Finally, Fig~.\ref{fig:final} highlights the  importance of considering the distribution of $R$ (rather than solely its mean) to understand the efficiency of any given measure. Here, the projected trajectories  with mean-shrinkage measures are displayed for both California  and Texas (for the period from 09/15 to 10/14) --- two states with similar median $R$ values (Fig.\ref{fig:rhist}), but different spreads ( see Table \ref{tab:table_quant}).  While shrinking the mean by 25\% is efficient in California reducing the epidemic, it is insufficient for Texas.
To conclude this section, these experiments highlight two potential interesting facts: (a) the efficiency of the measures is contingent on the distribution of the $R$, and not only its mean, and (b) measures targeting the tails might have comparable average accuracy (Fig.~.\ref{fig:France3}A), whilst better worst-case scenarios (Fig.~\ref{fig:France3}B) than measures targeted at the mean. As such, the heterogeneity of the distribution of $R$ appears to take on a significant importance in the type and scope of measures to be taken in order to control the epidemic.

\begin{table}[]
\resizebox{\columnwidth}{!}{%
    \centering
    \begin{tabular}{|c|c|c|c|c|c|c|} \hline
         Country&  60th Q. Cap & 0.6 $\times$ Mean & 90th Q. Cap & 0.9 $\times$ Mean & 95th Q. Cap & 0.95 $\times$ Mean  \\ \hline \hline
        France & 99.0\% &  99.3\% & 78.5\%  & 66.5\% &  58.4\% & 40.8 \% \\ \hline
        Germany & 99.1\% &  98.6\% & 58.7\%  & 44.8\% &  34.2\% & 21.5 \% \\ \hline
        UK & 92.6\% &  99.3\% & 55.8\%  & 66.6\% &  35.6\% & 40. \% \\ \hline
        Italy & 97.6\% &  99.4\% & 68.3\%  & 66.7\% &  48.5\% & 40.3 \% \\ \hline
        Spain & 94.3\% &  97.6\% & 21.8\%  & 27.7\% &  10.2\% & 11.9 \% \\ \hline
        Estonia & 94.9\% &  93.6\% & 80.4\%  & 61.2\% &  62.5\% & 34.4 \% \\ \hline
        Sweden & 87.5\% &  99.0\% & 49.9\%  & 64.1\% &  31.1\% & 41.3 \% \\ \hline
        Russia & 89.0\% &  99.5\% & 52.8\%  & 70.7\% &  34.8\% & 47.0 \% \\ \hline
        South Korea & 38.7\% &  37.9\% & 29.2\%  & 23.8\% &  21.9\% & 14.2 \% \\ \hline
        Mexico & 93.6\% &  98.8\% & 57.8\%  & 65.4\% &  37.2\% & 39.6 \% \\ \hline
        Colombia & 83.3\% &  97.0\% & 43.9\%  & 62.8\% &  29.7\% & 40.6 \% \\ \hline
        Texas & 97.4\% &  99.3\% & 70.1\%  & 72.8\% &  51.7\% & 45.8 \% \\ \hline
        Florida & 97.8\% &  99.0\% & 73.0\%  & 67.7\% &  53.0\% & 40.4 \% \\ \hline
        California & 94.7\% &  99.2\% & 61.6\%  & 67.4\% &  38.7\% & 42.2 \% \\ \hline
    \end{tabular}
    }
    \caption{Reduction in the final predicted incidence numbers of the first period (taken at day $t=30)$ achieved by different measures compared to predicted baseline (no measures) scenarios: $r = 1- \frac{P_{30}}{P^{(0)}_{30}}$. Comparison of the effect of capping at the $\{60, 90, 95\}$th quantiles, versus reducing the mean of the distribution to $\{60, 90, 95\}$\% of its original value.}
    \label{tab:final}
\end{table}

\section{Conclusion and Discussion}\label{sec:conclusion}

In conclusion, we have presented here an analysis targeted at assessing the impact of the heterogeneity of the reproductive number in predictive epidemic scenarios. In particular, we have shown that the modeling of this heterogeneity is crucial to correctly model extreme scenarios and characterize their uncertainty. Indeed,  using a Bayesian model, we have shown that the added variability is necessary to (a) provide better coverage of the confidence intervals, and thus, more appropriately quantify the uncertainty associated to a certain prediction or the effect of a given policy and (b) explain rare events and understand the formation of outbreaks --- which averaged models would not allow and which are nonetheless  crucial elements to take into account when weighting different scenarios.

Our analysis of the real data has also shown that considering the reproductive number as a distribution allows us to draw interesting nuances and contrasts between countries and stages of the epidemic, enabling to better capture the mean and variability of the reproductive number in the population. This analysis also hints to the importance and efficiency of capping the upper tails of the reproductive number. We emphasize again that our study does not aspire to draw predictive scenarios, but rather to understand how models and predictive scenarios are truly impacted by the choice and inherent variability of the $R$ -- and the great variability that we have imputed seems to highlight the need for a fine-grain analysis. \\


 \xhdr{Discussion} We now discuss the assumptions and possibilities of enhancement of our approach. In particular, the model that we propose in this paper makes three simplifying assumptions: (a) the number of cases is accurately measured (or rather, the under-ascertainment bias does not hinder our evaluation of the $R$), (b) the model is fitted per country, without considering neither the effect nor the similarities with neighboring regions and (c) the underlying distribution of $R$ is stationary.\\
\textit{ (a) Ascertainment bias:} Under-ascertainment has been a severe issue throughout the pandemic, as it mainly prevents a correct estimation of the Case (or Infection) Fatality Ratio (IFR/CFR). This under-ascertainment bias was especially pronounced in the early days of the pandemic, when testing needs completely outstripped availability as countries across the world were faced with unprecedented testing demand. From the mathematical perspective, denoting by $Y$ the true number of cases and $X$ the observed number of cases, the existence of under-ascertainment bias can be modelled as: $Y_t = \alpha_t X_t$, where $\alpha_t$ is a multiplicative constant greater than one which has to be estimated. The estimation of $\alpha_t$  is a difficult task: this bias is non-stationary, depends on testing availability and policies, and its evaluation is typically based on the evaluation of the number of deaths, with a lag of several weeks which also has to be estimated. However, assuming temporally locally stationary $\alpha_t$ (which we believe holds in the short time frames that we are considering), the relationship defining the reproductive number, $Y_{t+1} = \sum_s R Y_{t-s}$ , can also be written in terms of the observed cases $X_{t+1} = \sum_s R X_{t-s}$, and the impact of the under-ascertainment bias  on the estimation of the reproductive number is thus not as great nor crucial as on the Case-Fatality Ratio. While our model currently thus does not include the estimation of this under-ascertainment bias (consistently with most of the literature on the reproductive number $R$), it can thus be easily introduced as part of our pipeline, either as part of a pre-processing step in order to recover the true $Y$s, or as part of the Bayesian workflow, by modelling the coefficient $\alpha_t$ and associating it with a prior. Over the past few months, a number of papers have been dedicated to the issue \cite{omori2020ascertainment,noh2021estimation}, including some Bayesian approaches \cite{russell2020reconstructing}. Thus, while we focus here on the effect of the heterogeneity on the reproductive number $R$, we refer the reader to these works for a more in-depth discussion of the estimation of the under-ascertainment.
 \\
 \textit{(b)Spatial independence. } In this model, we fit each region independently, and do not consider the effect or similarity of neighbours. This simplifying assumption stems from the fact that, (i) mobility across countries was greatly reduced throughout the pandemic thus allowing us to consider countries as self-contained, and (ii) in view of the vastly diverging responses to the pandemic, we expect local policies and responses to the virus to create more variability in the reproductive number R than potential similarities in climate, demographic variables and social mores. The example of Sweden, who adopted a radically different course of action compared to neighbouring Norway and Finland, is a case in point.  However, further extensions of the model could include adding another hierarchical level to account for similarities between closely related countries (e.g, countries within the European Union, etc.) or by adding country-level  (temperature, population density, etc)  or population-level (data on employment type --- e.g, agricultural vs office-based---, average size/ age spread of the households, etc.) covariates $Z$. These variables have  indeed  been used to explain and account for heterogeneity in infection or transmission risk \cite{notari2020covid,notari2021temperature,liu2020impact,donnat2020bayesian,martin2020socio}, and could be added to further account for some of the heterogeneity of the reproductive number $R$, so that:
 $ R_t \sim f(Z)$.\\
 \textit{(c)} Stationarity. Finally, our model assumes a stationary $R$. This assumption is a valid approximation when fitting the model on short windows of time (i.e four weeks), in which public policies are not expected to change dramatically, or, should they change, the effect of the policy on the $R$ is not expected to have a huge impact. However, further extensions of the model could include allowing the distribution of $R$ to vary as a function of time, or allow some auto-regressive structure.

\vspace{2cm}
\xhdr{Disclaimer}
This model is a tool for exploring the effect of uncertainties and variation in the reproductive number $R$ for the virus and the effect of this variability in different types of interventions, but we do not claim to be predictive of disease dynamics for any specific populations (credit to McGee et al. for disclaimer).



\section{Bibliography}

{ \small
	\bibliographystyle{abbrv}
	\renewcommand{\refname}{\vspace{-5mm}}
	\bibliography{refs}

\begin{thebibliography}{10}

\bibitem{r0}
{\em R0: Estimation of R0 and Real-Time Reproduction Number from Epidemics,
  author = {Pierre-Yves Boelle, Thomas Obadia}, organization =
  {https://CRAN.R-project.org/package=R0}, address = {}, year = {2015}, url =
  {https://CRAN.R-project.org/package=R0},}.

\bibitem{akbarpour2020socioeconomic}
M.~Akbarpour, C.~Cook, A.~Marzuoli, S.~Mongey, A.~Nagaraj, M.~Saccarola,
  P.~Tebaldi, S.~Vasserman, and H.~Yang.
\newblock Socioeconomic network heterogeneity and pandemic policy response.
\newblock {\em University of Chicago, Becker Friedman Institute for Economics
  Working Paper}, (2020-75), 2020.

\bibitem{betancourt2017conceptual}
M.~Betancourt.
\newblock A conceptual introduction to hamiltonian monte carlo.
\newblock {\em arXiv preprint arXiv:1701.02434}, 2017.

\bibitem{betancourt2015hamiltonian}
M.~Betancourt and M.~Girolami.
\newblock Hamiltonian monte carlo for hierarchical models.
\newblock {\em Current trends in Bayesian methodology with applications},
  79(30):2--4, 2015.

\bibitem{carpenter2017stan}
B.~Carpenter, A.~Gelman, M.~D. Hoffman, D.~Lee, B.~Goodrich, M.~Betancourt,
  M.~Brubaker, J.~Guo, P.~Li, and A.~Riddell.
\newblock Stan: A probabilistic programming language.
\newblock {\em Journal of statistical software}, 76(1), 2017.

\bibitem{cave2020covid}
E.~Cave.
\newblock Covid-19 super-spreaders: Definitional quandaries and implications.
\newblock {\em Asian Bioethics Review}, page~1, 2020.

\bibitem{chang2020modelling}
S.~L. Chang, N.~Harding, C.~Zachreson, O.~M. Cliff, and M.~Prokopenko.
\newblock Modelling transmission and control of the covid-19 pandemic in
  australia.
\newblock {\em arXiv preprint arXiv:2003.10218}, 2020.

\bibitem{chatterjee2020healthcare}
K.~Chatterjee, K.~Chatterjee, A.~Kumar, and S.~Shankar.
\newblock Healthcare impact of covid-19 epidemic in india: A stochastic
  mathematical model.
\newblock {\em Medical Journal Armed Forces India}, 2020.

\bibitem{cori2013new}
A.~Cori, N.~M. Ferguson, C.~Fraser, and S.~Cauchemez.
\newblock A new framework and software to estimate time-varying reproduction
  numbers during epidemics.
\newblock {\em American journal of epidemiology}, 178(9):1505--1512, 2013.

\bibitem{daley2001epidemic}
D.~J. Daley and J.~Gani.
\newblock {\em Epidemic modelling: an introduction}, volume~15.
\newblock Cambridge University Press, 2001.

\bibitem{deforche2020age}
K.~Deforche.
\newblock An age-structured epidemiological model of the belgian covid-19
  epidemic.
\newblock {\em medRxiv}, 2020.

\bibitem{delamater2019complexity}
P.~L. Delamater, E.~J. Street, T.~F. Leslie, Y.~T. Yang, and K.~H. Jacobsen.
\newblock Complexity of the basic reproduction number (r0).
\newblock {\em Emerging infectious diseases}, 25(1):1, 2019.

\bibitem{dolbeault2020heterogeneous}
J.~Dolbeault and G.~Turinici.
\newblock Heterogeneous social interactions and the covid-19 lockdown outcome
  in a multi-group seir model.
\newblock {\em arXiv preprint arXiv:2005.00049}, 2020.

\bibitem{donnat2020bayesian}
C.~Donnat, N.~Miolane, F.~Bunbury, and J.~Kreindler.
\newblock A bayesian hierarchical network for combining heterogeneous data
  sources in medical diagnoses.
\newblock In {\em Machine Learning for Health}, pages 53--84. PMLR, 2020.

\bibitem{fraser2007estimating}
C.~Fraser.
\newblock Estimating individual and household reproduction numbers in an
  emerging epidemic.
\newblock {\em PloS one}, 2(8), 2007.

\bibitem{gelman2013bayesian}
A.~Gelman, J.~B. Carlin, H.~S. Stern, D.~B. Dunson, A.~Vehtari, and D.~B.
  Rubin.
\newblock {\em Bayesian data analysis}.
\newblock CRC press, 2013.

\bibitem{gomez2020mapping}
A.~G{\'o}mez-Carballa, X.~Bello, J.~Pardo-Seco, F.~Martin{\'o}n-Torres, and
  A.~Salas.
\newblock Mapping genome variation of sars-cov-2 worldwide highlights the
  impact of covid-19 super-spreaders.
\newblock {\em Genome Research}, 30(10):1434--1448, 2020.

\bibitem{grant2020dynamics}
A.~Grant.
\newblock Dynamics of covid-19 epidemics: Seir models underestimate peak
  infection rates and overestimate epidemic duration.
\newblock {\em medRxiv}, 2020.

\bibitem{he2020seir}
S.~He, Y.~Peng, and K.~Sun.
\newblock Seir modeling of the covid-19 and its dynamics.
\newblock {\em Nonlinear Dynamics}, pages 1--14, 2020.

\bibitem{hethcote2000mathematics}
H.~W. Hethcote.
\newblock The mathematics of infectious diseases.
\newblock {\em SIAM review}, 42(4):599--653, 2000.

\bibitem{hoffman2014no}
M.~D. Hoffman and A.~Gelman.
\newblock The no-u-turn sampler: adaptively setting path lengths in hamiltonian
  monte carlo.
\newblock {\em J. Mach. Learn. Res.}, 15(1):1593--1623, 2014.

\bibitem{hu2020identification}
K.~Hu, Y.~Zhao, M.~Wang, Q.~Zeng, X.~Wang, M.~Wang, Z.~Zheng, X.~Li, Y.~Zhang,
  T.~Wang, et~al.
\newblock Identification of a super-spreading chain of transmission associated
  with covid-19.
\newblock {\em medRxiv}, 2020.

\bibitem{kai2020universal}
D.~Kai, G.-P. Goldstein, A.~Morgunov, V.~Nangalia, and A.~Rotkirch.
\newblock Universal masking is urgent in the covid-19 pandemic: Seir and agent
  based models, empirical validation, policy recommendations.
\newblock {\em arXiv preprint arXiv:2004.13553}, 2020.

\bibitem{kermack1927contribution}
W.~O. Kermack and A.~G. McKendrick.
\newblock A contribution to the mathematical theory of epidemics.
\newblock {\em Proceedings of the royal society of london. Series A, Containing
  papers of a mathematical and physical character}, 115(772):700--721, 1927.

\bibitem{lekone2006statistical}
P.~E. Lekone and B.~F. Finkenst{\"a}dt.
\newblock Statistical inference in a stochastic epidemic seir model with
  control intervention: Ebola as a case study.
\newblock {\em Biometrics}, 62(4):1170--1177, 2006.

\bibitem{liu2020impact}
J.~Liu, J.~Zhou, J.~Yao, X.~Zhang, L.~Li, X.~Xu, X.~He, B.~Wang, S.~Fu, T.~Niu,
  et~al.
\newblock Impact of meteorological factors on the covid-19 transmission: A
  multi-city study in china.
\newblock {\em Science of the total environment}, 726:138513, 2020.

\bibitem{lyra2020covid}
W.~Lyra, J.~D. do~Nascimento, J.~Belkhiria, L.~de~Almeida, P.~P. Chrispim, and
  I.~de~Andrade.
\newblock Covid-19 pandemics modeling with seir (+ caqh), social distancing,
  and age stratification. the effect of vertical confinement and release in
  brazil.
\newblock {\em medRxiv}, 2020.

\bibitem{mackenzie}
D.~Mackenzie.
\newblock Why coronavirus superspreaders may mean we avoid a deadly pandemic.
\newblock {\em New Scientist}.

\bibitem{martin2020socio}
C.~A. Martin, D.~R. Jenkins, J.~S. Minhas, L.~J. Gray, J.~Tang, C.~Williams,
  S.~Sze, D.~Pan, W.~Jones, R.~Verma, et~al.
\newblock Socio-demographic heterogeneity in the prevalence of covid-19 during
  lockdown is associated with ethnicity and household size: Results from an
  observational cohort study.
\newblock {\em EClinicalMedicine}, 25:100466, 2020.

\bibitem{noh2021estimation}
J.~Noh and G.~Danuser.
\newblock Estimation of the fraction of covid-19 infected people in us states
  and countries worldwide.
\newblock {\em PloS one}, 16(2):e0246772, 2021.

\bibitem{notari2021temperature}
A.~Notari.
\newblock Temperature dependence of covid-19 transmission.
\newblock {\em Science of The Total Environment}, 763:144390, 2021.

\bibitem{notari2020covid}
A.~Notari and G.~Torrieri.
\newblock Covid-19 transmission risk factors.
\newblock {\em arXiv preprint arXiv:2005.03651}, 2020.

\bibitem{omori2020ascertainment}
R.~Omori, K.~Mizumoto, and H.~Nishiura.
\newblock Ascertainment rate of novel coronavirus disease (covid-19) in japan.
\newblock {\em International Journal of Infectious Diseases}, 96:673--675,
  2020.

\bibitem{pandey2020seir}
G.~Pandey, P.~Chaudhary, R.~Gupta, and S.~Pal.
\newblock Seir and regression model based covid-19 outbreak predictions in
  india.
\newblock {\em arXiv preprint arXiv:2004.00958}, 2020.

\bibitem{rai2020estimates}
B.~Rai, A.~Shukla, and L.~K. Dwivedi.
\newblock Estimates of serial interval for covid-19: A systematic review and
  meta-analysis.
\newblock {\em Clinical epidemiology and global health}, 2020.

\bibitem{read2020novel}
J.~M. Read, J.~R. Bridgen, D.~A. Cummings, A.~Ho, and C.~P. Jewell.
\newblock Novel coronavirus 2019-ncov: early estimation of epidemiological
  parameters and epidemic predictions.
\newblock {\em medRxiv}, 2020.

\bibitem{read2020choir}
R.~Read.
\newblock A choir decided to go ahead with rehearsal. now dozens of members
  have covid-19 and two are dead.
\newblock {\em Los Angeles Times}, 29, 2020.

\bibitem{rockett2020revealing}
R.~J. Rockett, A.~Arnott, C.~Lam, R.~Sadsad, V.~Timms, K.-A. Gray, J.-S. Eden,
  S.~Chang, M.~Gall, J.~Draper, et~al.
\newblock Revealing covid-19 transmission in australia by sars-cov-2 genome
  sequencing and agent-based modeling.
\newblock {\em Nature medicine}, 26(9):1398--1404, 2020.

\bibitem{russell2020reconstructing}
T.~W. Russell, N.~Golding, J.~Hellewell, S.~Abbott, L.~Wright, C.~A. Pearson,
  K.~van Zandvoort, C.~I. Jarvis, H.~Gibbs, Y.~Liu, et~al.
\newblock Reconstructing the early global dynamics of under-ascertained
  covid-19 cases and infections.
\newblock {\em BMC medicine}, 18(1):1--9, 2020.

\bibitem{silva2020covid}
P.~C. Silva, P.~V. Batista, H.~S. Lima, M.~A. Alves, F.~G. Guimar{\~a}es, and
  R.~C. Silva.
\newblock Covid-abs: An agent-based model of covid-19 epidemic to simulate
  health and economic effects of social distancing interventions.
\newblock {\em Chaos, Solitons \& Fractals}, 139:110088, 2020.

\bibitem{earlyR}
P.~N. Thibaut~Jombart, Anne~Cori.
\newblock {\em earlyR: Estimation of Transmissibility in the Early Stages of a
  Disease Outbreak}.
\newblock https://CRAN.R-project.org/package=earlyR, 2017.

\bibitem{wu2020nowcasting}
J.~T. Wu, K.~Leung, and G.~M. Leung.
\newblock Nowcasting and forecasting the potential domestic and international
  spread of the 2019-ncov outbreak originating in wuhan, china: a modelling
  study.
\newblock {\em The Lancet}, 2020.

\bibitem{zhang2020evaluating}
Y.~Zhang, Y.~Li, L.~Wang, M.~Li, and X.~Zhou.
\newblock Evaluating transmission heterogeneity and super-spreading event of
  covid-19 in a metropolis of china.
\newblock {\em International Journal of Environmental Research and Public
  Health}, 17(10):3705, 2020.

\bibitem{zhao2020modeling}
S.~Zhao and H.~Chen.
\newblock Modeling the epidemic dynamics and control of covid-19 outbreak in
  china.
\newblock {\em Quantitative Biology}, pages 1--9, 2020.

\bibitem{zhao2020preliminary}
S.~Zhao, Q.~Lin, J.~Ran, S.~S. Musa, G.~Yang, W.~Wang, Y.~Lou, D.~Gao, L.~Yang,
  D.~He, et~al.
\newblock Preliminary estimation of the basic reproduction number of novel
  coronavirus (2019-ncov) in china, from 2019 to 2020: A data-driven analysis
  in the early phase of the outbreak.
\newblock {\em International Journal of Infectious Diseases}, 2020.

\end{thebibliography}
}

\appendix
\section{Assessing Goodness of fit and Convergence Diagnostics}\label{appendix:dets}

\xhdr{COVID-19 Model Fitting: Convergence Diagnostics}

In this appendix, we discuss in greater details the choice of the model and convergence diagnostics

\xhdr{1. Choice of model} The first step of this discussion consists in the choice of the model made in Section \ref{sec:model}. As discussed in Section \ref{sec:sim}, this choice is important, since it determines the amount of variability of the distribution of $R$ --- which we characterize through the evaluation of its coefficient of variation. All of the models in this paper are based on a modification of the model by Cori et al \cite{cori2013new}, which assumes a Poisson distribution for the number of new incident cases:

$$ X_{t+1} \sim \text{Poisson}( \lambda_t )$$
where $\lambda_t =  \sum_{k=1}^K w_k R_t I_{t-k}$. As described in the main text, for any $s\geq 1$, $w_s$ is a vector such that $\sum_s w_s=1$. In other words, this equation models $I_t$ as a weighted sum of all the new secondary cases generated by each single new cases in the past $K$ days, and where the varying weights capture the varying levels of infectivity. Indeed, infected cases are not as likely to transmit the disease on day 0 of their infection as on day 8 for instance. 

Having chosen a parametrisation for the number of new cases per day, it remains necessary to choose a set of priors for the different parameters. As detailed in Section   \ref{sec:sim}, we  compared the following four different models, which correspond to different behaviours of the coefficient of variation:\\

\xhdr{Model $M_1$} $I_t \sim \text{Gamma}(\alpha X_t, \beta  )$
This is assuming each $R_i$ is sampled from an independent gamma, so that  $I_t \overset{\Delta}{=} \sum_{i=1}^{X_t} \text{Gamma}(\alpha, \beta  )\overset{\Delta}{=}\text{Gamma}(\alpha X_t, \beta  )$. In other words, this is akin to the additive model considered in Section \ref{sec:sim}.

\xhdr{Model $M_2$} $I_t  \sim \text{Gamma}(\alpha X_t, \beta/X_t )$
This assumes that we have a multiplication factor in front of the number of cases. As discussed in Section  \ref{sec:model} of the main paper, this might be more amenable to accounting for under reporting.

\xhdr{Model $M_3$} $I_t  \sim \text{Gamma}(\alpha X_t \sqrt{I_t}, \beta/\sqrt{I_t} )$
This model is in a similar vein as the previous,  but allows less variance and has a smaller coefficient of variation. We have fitted here using two versions of the $w$ parameter, one similar to the approach suggested by Cori et al \cite{cori2013new}, using the serial interval for COVID-19, and another ($W=w_1$) using the value of $w$ suggested in the main text.

\xhdr{Model  $M_4$} $I_t  \sim \text{Gamma}(\alpha X_t \sqrt{I_t}, \beta/\sqrt{I_t} ), \quad w \sim \text{Dirichlet}(w_0)$

\begin{figure}
    \centering
    \includegraphics[width=\textwidth]{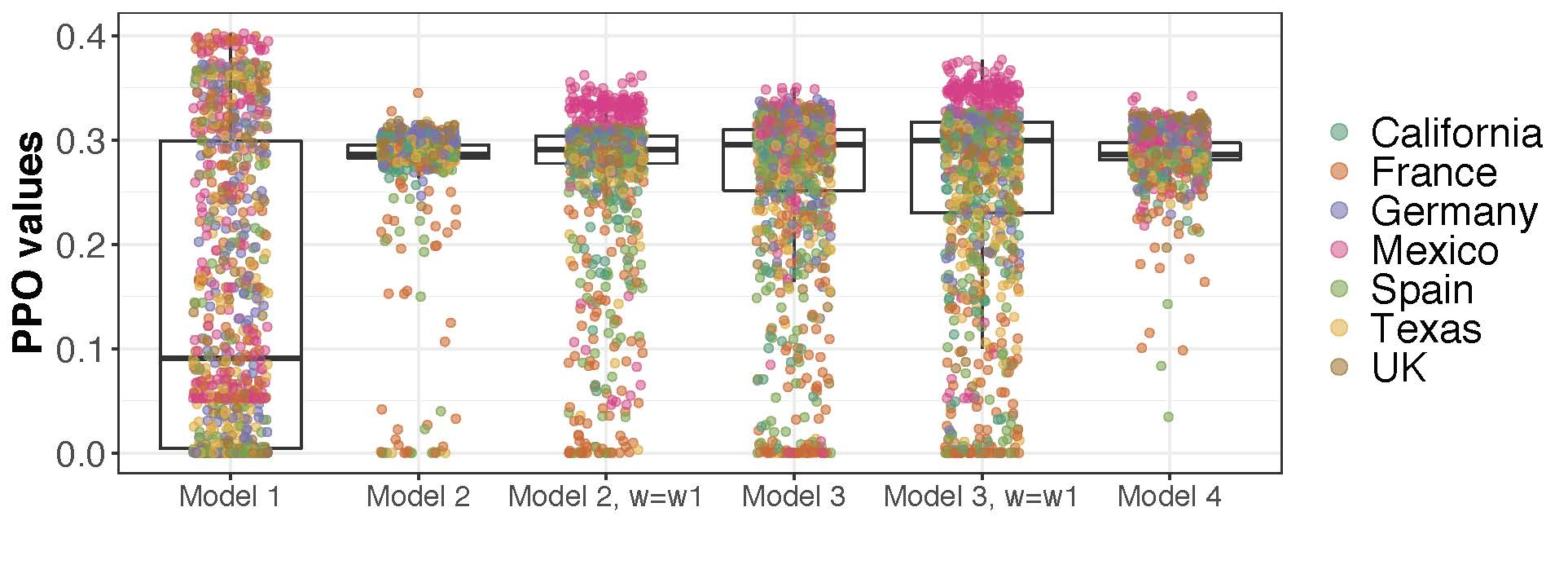}
  {\centering  \caption{Posterior Predictive Ordinates for the different models, coloured by country.}}
    \label{fig:PPO}
\end{figure}

To assess goodness of fit, we fitted all four models on a subset of 7 countries or regions, randomly chosen by the authors and in order to capture different behaviours of the pandemic:  France, the United Kingdom, Spain, Germany, California, Texas, and Mexico. We also compared the impact of choosing a value $w$ to be $w\propto \{1:K\}$ (as suggested in the main text) or the $w_1$ based on the infectious profile described in Cori et al \cite{cori2013new}. The comparison was done using criteria used in the Bayesian literature for model comparison, as suggested in Gelman et al \cite{gelman2013bayesian}. We investigated here three of these criteria.\\
\textit{(1)Posterior predictive ordinate (PPO) }. The Posterior Predictive Ordinate is the density of the posterior predictive distribution evaluated at observation $y_i$.  This quantity can be used
to estimate the probability of observing $y_i$ given $\mathbf{y}$:
$$ \text{PPO}_i = f(y_i|y) = \int f(y_i|\theta) f(\theta|y) d\theta$$
It can be estimated through the following formula:
$$ \widehat{\text{PPO}_i }= \frac{1}{S} \sum_{i=1}^S f(y_i |\theta^{(s)}).$$
In this case, the higher the PPO, the better the fit of the model. Results are shown in Fig.~\ref{fig:PPO}. These PPOs can then be averaged to compute the log pointwise predictive density (LPPD)
$$ \text{LPPD} =  \sum_{i=1}^n \log(\widehat{\text{PPO}_i }) =  \sum_{i=1}^n \log( \frac{1}{S} \sum_{i=1}^S f(y_i |\theta^{(s)}) ) .$$
\textit{(2) Conditional Predictive Ordinate}. As pointed out in Gelman et al \cite{gelman2013bayesian},, the LPPD is an overestimate of the expected  log-pointwise predictive density for future data -- we thus have to correct for its bias. The conditional predictive ordinate (CPO) is based on leave-one-out-cross-validation and partly answer this question. CPO estimates the probability of observing $y_i$
in the future if after
having already observed $y_{-i}$.Low CPO values suggest possible outliers, high-leverage and
influential observations:
$$\text{CPO}_i = f(y| y_{-i}) =\Big[ \int\frac{1}{f(y_i|\theta)} f(\theta|y) d\theta \Big]^{-1}$$.

\begin{figure}
    \centering
    \includegraphics[width=\textwidth]{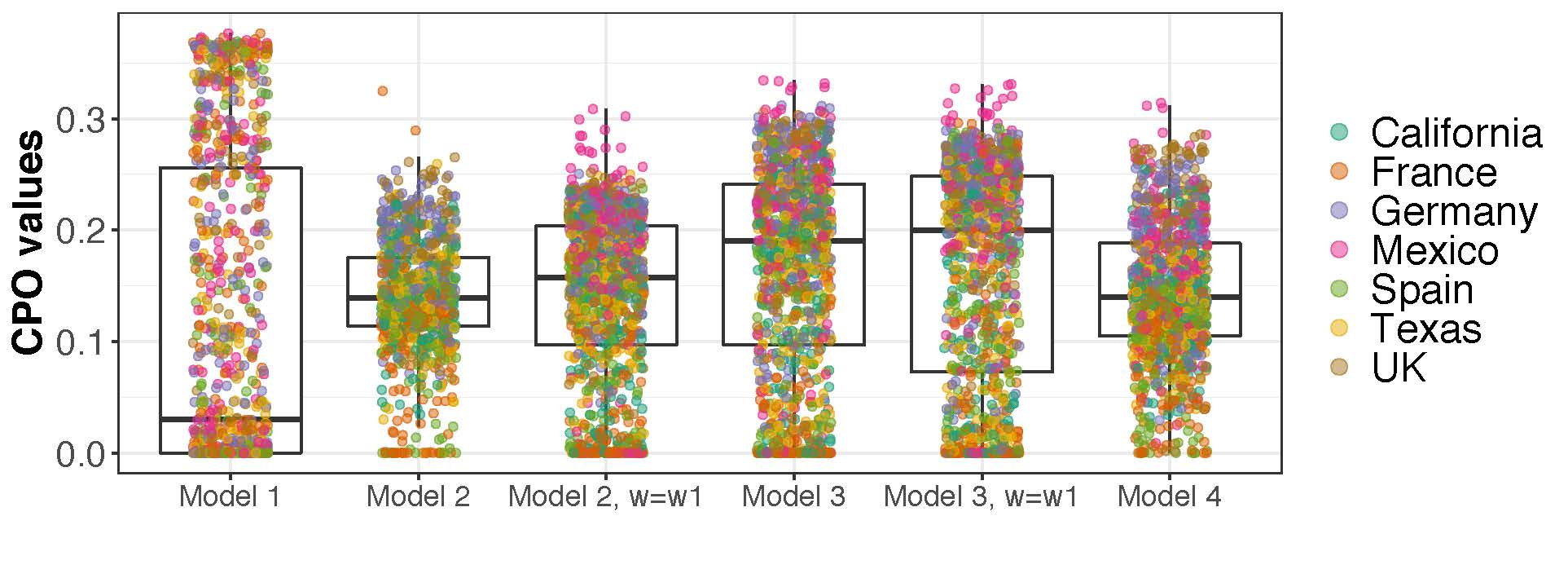}
    \caption{Conditional Predictive Ordinates for the different models, coloured by country.}
    \label{fig:CPO}
\end{figure}

The CPO can be estimated by taking the inverse of the posterior
mean of the inverse density function value of $y_i$ (harmonic mean of the
likelihood of $y_i$). Thus:
$$ \widehat{\text{CPO}_i }= \Big[\frac{1}{S} \sum_{i=1}^S \frac{1}{f(y_i |\theta^{(s)})}\Big]^{-1}.$$
Results are shown in Fig.~\ref{fig:CPO}. The individual CPOs can then be summarised to compare models using the  log-pseudo marginal likelihood (LPML) . 
The best model amongst competing models have the largest
LPML.
$$ \text{LPML} =\frac{1}{n} \sum_{i=1}^n \log(CPO_i)$$
A comparison of the values of the LPPD and LPML is provided in Table~\ref{tab:gof}.\\
\textit{(3)Likelihood of the model} Another model evaluation metric consists of the likelihood of the observations in the fitted model.  Model comparison can then be achieved by looking at the Bayes Ratios, similarly to as in hypothesis testing. As underlined in Gelman et al \cite{gelman2013bayesian}, this ratio is not necessarily the most telling evaluation of the goodness of fit, since it depends on a number of untestable assumptions, and assumes that one of the models is correct. We chose to include here for the sake of completeness and as  a sanity check.\\

As we can see, the models allowing for varying $R$ through time (CV coefficient not constant, ie, Models $M_2$ and $M_3$) perform better. We choose Model $M_2$ due to its simplicity for similar levels of performance on the LPML and LPPD compared to Model $M_4$ and marginal superiority over the others on these two metrics in particular. We note though that models $M_2$ (with the Cory $w=w_1$ and the linear $w$ ) perform roughly similarly. Due to its better LPML value, we preferred adopting model $M_2$.

\begin{table}[]
    \centering
    \begin{tabular}{|c|c|c|c|} \hline
    Model &  Average log-likelihood & Log Pointwise &  Log-Pseudo  \\
         &   & Predictive Density&  Marginal Likelihood \\
    \hline \hline
        Model M1 &  -19.4 & -4.98 & -Inf \\  \hline
       Model M2  & -2.02 & -1.65 &-2.73 \\  \hline
        Model M2, $w=w_1$  & -2.32 &-1.22& -3.68 \\  \hline
       Model M3  & -2.87 & -2.13  & -3.71  \\  \hline
        Model M3, $w=w_1$ & -3.18 &  -2.15 & -4.32 \\  \hline
                Model M4  & -1.43 & -1.24 &  -2.28 \\  \hline
    \end{tabular}
    \caption{Goodness of Fit Metrics for the different models}
    \label{tab:gof}
\end{table}

\xhdr{2. Numerical Procedure}

We fit the model using Hamiltonian Monte Carlo (No U Turn Samples) with the {\tt R} package the RStan \cite{carpenter2017stan}. We use 10 chains, with 5,000 warmup iterations and 1,000 sampling steps. Using these 10,000 posterior samples, we estimated the posterior median of the posterior and 95\% credible interval (CrI) for each time point.

We assessed convergence of the chains using the Gelman-Rubin convergence diagnostic, that is, to assume convergence to the posterior, we ensured that $\hat{R} \leq 1.1$  across all chains and that the number of effective samples was more than 20\% of the number of samples.  We also checked the mixing of all chains. The following plots show convergence diagnostics for all the countries that were selected for the purpose of this analysis.

\begin{figure}
    \centering
    \includegraphics[width=\textwidth]{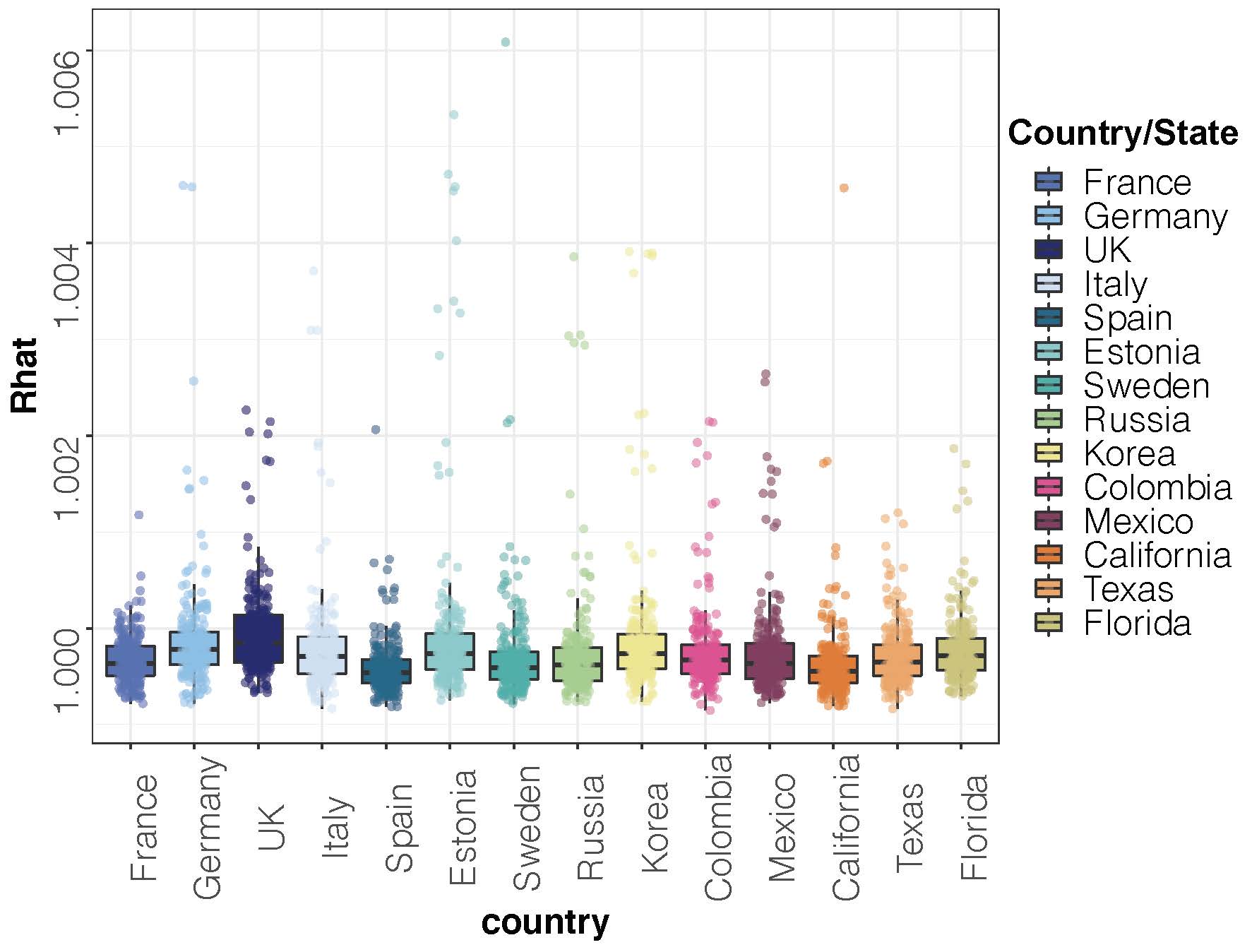}
    \caption{$\hat{R}$ for the different parameters in the model, across groups.}
\end{figure}

\begin{figure}
    \centering
    \includegraphics[width=\textwidth]{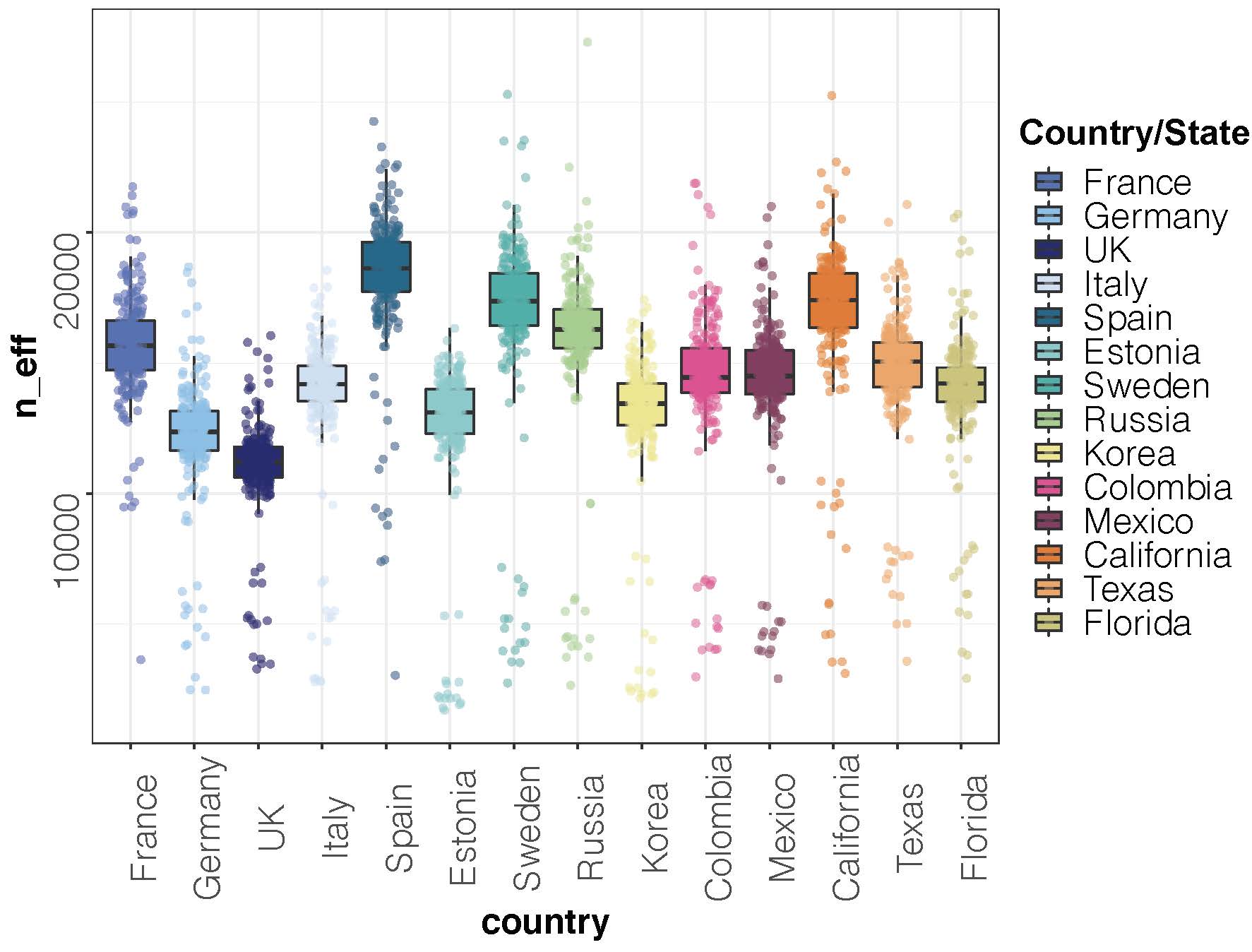}
        \caption{$n_{\text{eff}}$ for the different parameters in the model, across groups.}\label{fig:covdiagnostics:b}
\end{figure}

\section{Data}\label{appendix:dets_data}

The data that we used here consists in the new incident cases (as per the JHU \footnote{The data is publicly available at the following  \href{link}{https://github.com/CSSEGISandData/COVID-19}.} dataset). We use country (and in the case of the US, county) level data to fit the different $R$s. The data required a little bit of pre-processing, including:
\begin{itemize}
    \item Converting the cumulative counts to daily incidence counts
    \item Thresholding to 0 the negative entries
    \item Using the 7-day rolling average counts to even out potential weekly (and weekend effects) observable in the raw data.
\end{itemize}


\onecolumn
\newpage



\end{document}